\documentclass[%
reprint,
superscriptaddress,
 amsmath,amssymb,
aps,
prb,
]{revtex4-2}

\usepackage{pifont}
\usepackage{verbatim}
\usepackage{subfigure}
\usepackage{graphicx}
\usepackage{dcolumn}
\usepackage{bm}
\usepackage{xcolor} 
\usepackage{hyperref}
\usepackage{url}

\begin{document}

\title{Filling anomaly for general 2D and 3D $C_4$ symmetric lattices}
%

\author{Yuan Fang}
\affiliation{Department of Physics and Astronomy, Stony Brook University, Stony Brook, New York 11974, USA}

\author{Jennifer Cano}
\affiliation{Department of Physics and Astronomy, Stony Brook University, Stony Brook, New York 11974, USA}
\affiliation{Center for Computational Quantum Physics, The Flatiron Institute, New York, New York 10010, USA}

\date{\today}

\begin{abstract}
In this manuscript, we derive symmetry indicator formulas for the filling anomaly on 2D square lattices with and without time reversal, inversion symmetry, or their product, in the presence of spin-orbit coupling.
We go beyond previous work by considering lattices with atoms occupying multiple Wyckoff positions. 
We also provide an algorithm using the Smith normal form that systematizes the derivation.
The formulas determine the corner charge in 2D atomic or fragile topological insulators, as well as in 3D insulators and semimetals by studying their 2D slices. 
We apply our results to a 3D tight-binding model on a body-centered tetragonal lattice, whose projection into the 2D plane has two atoms in the unit cell.
Our symmetry indicators correctly describe the higher-order hinge states and Fermi arcs in cases where the existing indicators do not apply. 
\end{abstract}

\maketitle

\section{Introduction\label{sec_intro}}


The discovery of higher order topological insulators (HOTIs) has refined the notion of the bulk-boundary correspondence \cite{benalcazar2017electric,schindler2018higher,song2017d,langbehn2017reflection,benalcazar2017quantized,geier2018second,trifunovic2019higher,schindler2018bismuth,khalaf2018symmetry,khalaf2018higher,ezawa2018minimal,ezawa2018higher,matsugatani2018connecting,imhof2018topolectrical,peterson2018quantized,serra2018observation,noh2018topological,you2018higher,queiroz2019partial,fang2020higher}.
Specifically, an order-$d$ topological insulator in $D$ dimensions exhibits gapless modes on $(D-d)$-dimensional surface, where $d=1$ corresponds to the usual bulk-boundary correspondence \cite{schnyder2008classification,schnyder2009classification,kitaev2009periodic}.

In this manuscript, we consider the case where $d=D$. An order-$D$ topological insulator in $D$ dimensions exhibits zero-dimensional corner-localized mid-gap states \cite{benalcazar2017quantized,benalcazar2017electric,benalcazar2019quantization,schindler2019fractional,watanabe2020fractional,hirayama2020higher}.
Unlike a Chern insulator \cite{panati2007triviality,brouder2007exponential}, $\mathbb{Z}_2$ topological insulator \cite{soluyanov2011wannier}, or topological crystalline insulator \cite{bradlyn2017topological,po2017symmetry}, an order-$D$ topological insulator in $D$ dimensions does not require an obstruction to the existence of symmetric, maximally localized Wannier functions.
Instead, the corner charge can result from an obstructed atomic limit (OAL) phase \cite{bradlyn2017topological}, where the bulk is a band insulator that permits maximally localized and symmetric Wannier functions, but such that the Wannier centers cannot be continuously deformed into the positions of the atoms without breaking symmetry or closing the bulk (or surface \cite{khalaf2019boundary}) band gap.
This mismatch between the bulk atoms and Wannier centers has been dubbed the filling anomaly \cite{benalcazar2019quantization}.
In a symmetric finite-sized system at charge neutrality with no polarization or surface states,
a filling anomaly results in a non-zero corner charge, quantized by crystal symmetry.

It is desirable to compute the filling anomaly and corner charge from bulk properties.
To this end, there have been two recent approaches.
The first, which applies to any lattice, regardless of symmetry, is to generalize the modern theory of polarization \cite{kingsmith1993theory,vanderbilt1993electric,resta1994macroscopic} by determining the corner charge from a bulk multipole moment \cite{PhysRevB.92.041102,wheeler2019many,kang2019many,ono2019difficulties,trifunovic2019geometric,trifunovic2020bulk,ren2020quadrupole,watanabe2020corner}.
The second, which is taken in this manuscript, is to develop a theory of symmetry indicators, i.e., formulas derived in terms of the symmetry representations of the Bloch wavefunctions at high-symmetry momenta.
Symmetry indicators have been very successful in classifying topological crystalline insulators \cite{bradlyn2017topological,vergniory2017graph,elcoro2017double,po2017symmetry,bradlyn2018band,cano2018building,khalaf2018symmetry,song2018quantitative,vergniory2019complete,zhang2019catalogue,cano2020band}
starting with the inversion eigenvalue formulas for 2D and 3D $\mathbb{Z}_2$ topological insulators \cite{fu2007topological}.

\begin{figure}[b]
    \centering
    \includegraphics[width=\linewidth]{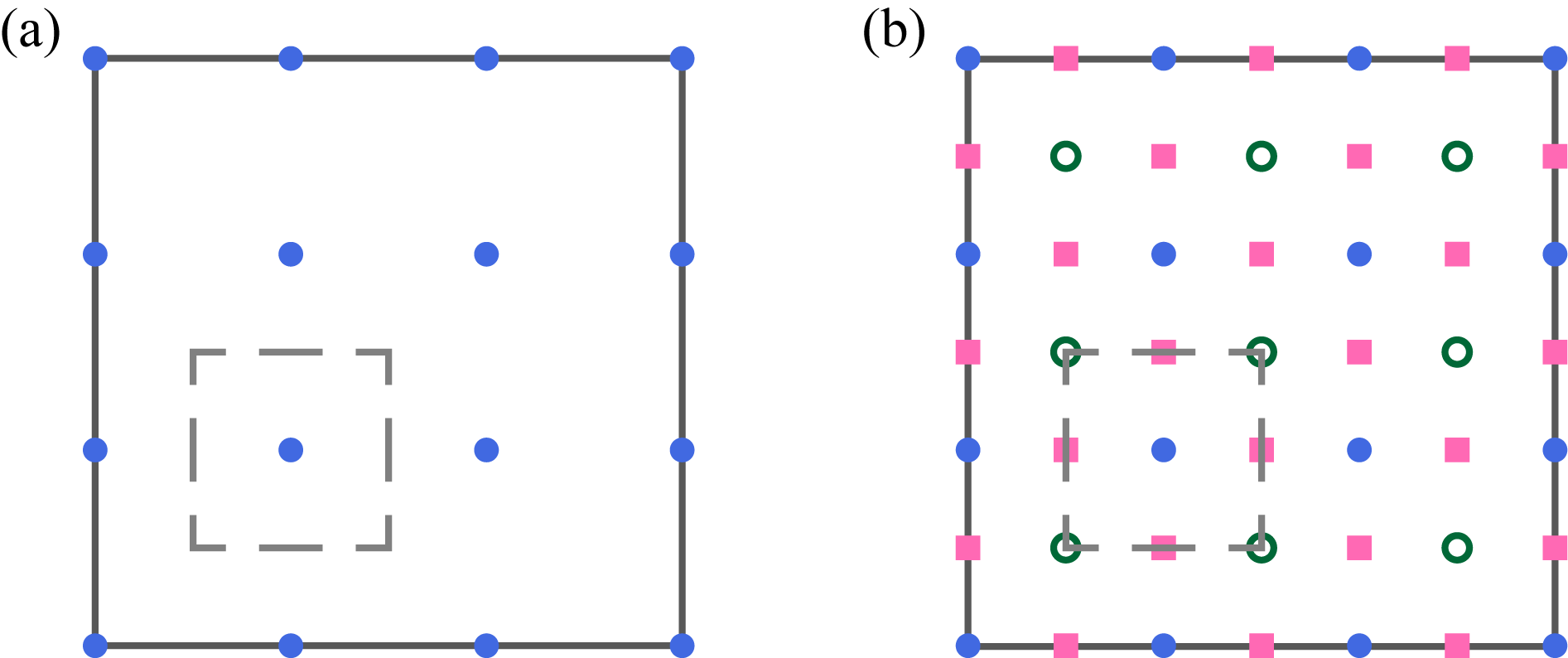}
    \caption{Finite symmetric square lattices with open boundary conditions shown for (a) the simple square lattice and (b) a lattice with atoms in multiple Wyckoff positions.
    Solid blue and hollow green dots and pink squares correspond to different Wyckoff positions on the square lattice. Dashed gray lines indicate the primitive unit cell. The large black square outlines the finite-sized system (atoms on the boundary are included in the finite-sized system). 
    In (b), the finite-sized system includes a fractional number of unit cells, as evidenced by the different number of solid and hollow dots, while the unit cell contains one of each. 
    }
    \label{fig_lattice}
\end{figure}

Recently, symmetry indicators have been derived for the filling anomaly and corner charge in OALs in certain 2D crystals~\cite{benalcazar2019quantization,schindler2019fractional}.
However, the results do not necessarily apply when there are multiple atoms in the unit cell.
Specifically, Refs.~\cite{benalcazar2019quantization,schindler2019fractional} limited their consideration to crystals for which there exists a symmetric finite-size termination containing an integer number of unit cells.
Such a termination does not exist for a crystal with atoms occupying multiple distinct maximal Wyckoff positions.
Instead, for such a crystal, all symmetric terminations contain a fractional number of unit cells, as shown in Fig.~\ref{fig_lattice}.

In this manuscript, we develop a method to compute symmetry indicators for the filling anomaly in the general case of a crystal with atoms occupying any number of Wyckoff positions.
The method has two steps:
we first compute the filling anomaly in terms of the number of Wannier functions centered at each Wyckoff position (which has also been done recently in Ref.~\cite{watanabe2020corner}).
Second, we compute the number of Wannier centers at each Wyckoff position in terms of symmetry indicators using elementary band representations.
The second step is accomplished via an algorithm that automates the calculation of symmetry indicators, introduced in this work.
We apply our method to the square lattice with and without time reversal, inversion and their product, corresponding to the layer groups $p4/m1'$, $p4/m'$, $p4$, $p4/m$, and $p41'$; the results are in Eqs.~(\ref{eqn_eta_mult}), (\ref{eqn_eta1}), (\ref{eqn_eta2}), (\ref{eqn_eta3}), and (\ref{eqn_eta4}), respectively.
Our results provide a necessary generalization of formulas in previous work \cite{benalcazar2019quantization,schindler2019fractional}, which can give an incorrect result when there are multiple atoms in the unit cell.

In addition to diagnosing band structures in 2D, the results of our work can be used to compute 2D invariants for slices of the 3D Brillouin zone, which is crucial to diagnosing the topology of 3D semimetals \cite{wieder2020strong,lin2018topological,ghorashi2020higher} and some 3D HOTIs \cite{song2017d}.
We present an example of this in Sec.~\ref{sec_3d}.

The manuscript is organized as follows. In Sec.~\ref{sec_2d}, we review the concept of the filling anomaly and its connection to corner charge. 
We derive relations between the filling anomaly and the number of Wannier centers at each Wyckoff position.
We then derive symmetry indicator formulas for the filling anomaly (and therefore corner charge) that apply to the square lattice.\par

In Sec.~\ref{sec_3d}, we build a 3D body-centered tetragonal (BCT) model. We analyze the HOTI phase and higher order semi-metal phase in this model by applying our formulas to each $k_z$ slice, which corresponds to a 2D system with two distinct atoms in the primitive unit cell.
We verify our new formulas by numerically calculating the corner/hinge states.
Our example demonstrates why the previous formulas in Refs.~\cite{benalcazar2019quantization,schindler2019fractional} do not hold for a BCT lattice. In Sec.~\ref{sec_discussion}, we summarize our results and discuss future directions.\par

\section{2D square lattice\label{sec_2d}}
We consider gapped 2D systems on the square lattice with no gapless edge modes and no bulk polarization. These systems are either (possibly obstructed) atomic limits or fragile topological phases, where all strong symmetry indicators vanish \cite{po2017symmetry}. 
In addition to the $\pi/2$ rotation symmetry of the square lattice, which we denote by $C_4$, 
we consider the presence of time reversal $\cal T$, inversion $\cal I$, and/or their product,  corresponding to the following 2D symmetry groups, known as layer groups: $p4/m1'$ ($C_4$, $\cal{T}$, $\cal{I}$); $p4/m'$ ($C_4$, $\cal{TI}$); $p4/m$ ($C_4$, $\cal I$); $p41'$ ($C_4$, $\cal T$); and $p4$ ($C_4$ only), where the symmetry operations in parenthesis indicate the generators, excluding translations.
The layer groups are listed in international notation, where $4$ indicates the $C_4$ rotation; $1'$ indicates $\cal{T}$, $/m$ indicates $\cal{I}$; and $m'$ indicates $\cal{IT}$ \cite{litvin}.
{The layer group $p41'$ does not have a complete symmetry indicator formula, as pointed out in Ref.~\cite{schindler2019fractional}; we derive a partial indicator in Sec.~\ref{sec_p41p}.}

In this section, we describe the method for deriving the symmetry indicator for the filling anomaly (Secs.~\ref{sec_corresp}, \ref{sec_2d_method} and \ref{sec_2d_method_symm}), which can be generalized to any crystal symmetry group in any dimension.
In Sec.~\ref{sec_2d_p4m1p}, we apply the method to the symmetry group $p4/m1'$ with spin-orbit coupling (SOC).
We first re-derive the formula for the case of only one atom in the unit cell \cite{schindler2019fractional} and then derive a new formula for the situation where atoms occupy multiple Wyckoff positions.
We generalize to the layer groups $p4/m'$, $p4$, $p4/m$, and $p41'$ with SOC in Sec.~\ref{sec_2d_gen} and summarize in Sec.~\ref{sec_summary_2D}.

\par

\subsection{\label{sec_corresp}Bulk-corner correspondence}

Topologically trivial bands have symmetric and exponentially localized Wannier functions \cite{bradlyn2017topological,po2017symmetry}.
When the Wannier centers cannot be continuously moved to coincide with the atom positions while obeying crystal symmetry, the system is in an OAL phase \cite{bradlyn2017topological}. Despite having exponentially localized Wannier centers, OALs are non-trivial in the sense that they are separated by a gap-closing phase transition from the trivial phase (where the Wannier centers coincide with the atomic positions). Canonical examples include the Su-Schrieffer-Heeger model in 1D \cite{PhysRevLett.42.1698} and the quadrupole insulator in 2D \cite{benalcazar2017quantized}.

Due to the mismatch between the atomic positions and Wannier centers, OALs can sometimes support mid-gap corner-localized states.
The connection between bulk Wannier centers and mid-gap corner charge is called the bulk-corner correspondence.
The existence of mid-gap corner states indicates that in a finite-sized system with open boundaries, the number of filled bulk valence states is different from the filling required for charge neutrality. Thus, an OAL with mid-gap corner states can either be neutral or symmetric, but not both.
This difference between the neutral and symmetric fillings is called the filling anomaly \cite{benalcazar2019quantization}.
A nontrivial filling anomaly requires not only that the number of filled states differs from the charge neutral filling, but also that the difference cannot be accounted for by adding or removing electrons to the boundary in a symmetry-preserving way.
Thus, the filling anomaly remains robust even if the mid-gap states are pushed up(down) in energy into the conduction(valence) bands by a boundary potential.

In the symmetry groups with time-reversal ($p4/m1'$ and $p41'$), the filling anomaly is defined mod $8$ because one can add or remove eight electrons to the boundary of a finite-sized system without breaking these symmetries (by adding a Kramers pair of time-reversed partners to the four corners of a square lattice). 
In the systems without time-reversal ($p4/m'$, $p4/m$, or $p4$), the filling anomaly is defined mod $4$ because one can add four electrons to the boundary of a finite-sized system without breaking these symmetries, as electrons need not come in Kramers pairs.
\par

In this work, we are interested in the filling anomaly that results from purely corner charge.
Therefore, we limit ourselves to systems without gapless surface states, which excludes systems with a bulk polarization or a nontrivial topological invariant.

To compute the corner charge, we note that the symmetrically terminated square lattice with filling anomaly $\eta$ has net charge $\eta e$. 
Symmetry requires the charge $\eta e$ is symmetrically sitting at the four corners, resulting in a corner charge $Q_c$ on the square lattice  \cite{benalcazar2019quantization},
\begin{equation}
\label{eqn_qc}
    Q_c=\frac{\eta }{4}e.
\end{equation}
Since, as discussed above, $\eta$ is defined either mod 8 or mod 4, it follows that $Q_c$ is defined mod $2e$ with time-reversal and mod $e$ without. 
\par

\subsection{\label{sec_2d_method}Defining the filling anomaly}

\begin{figure}[b]
    \centering
    \includegraphics[width=0.85\linewidth]{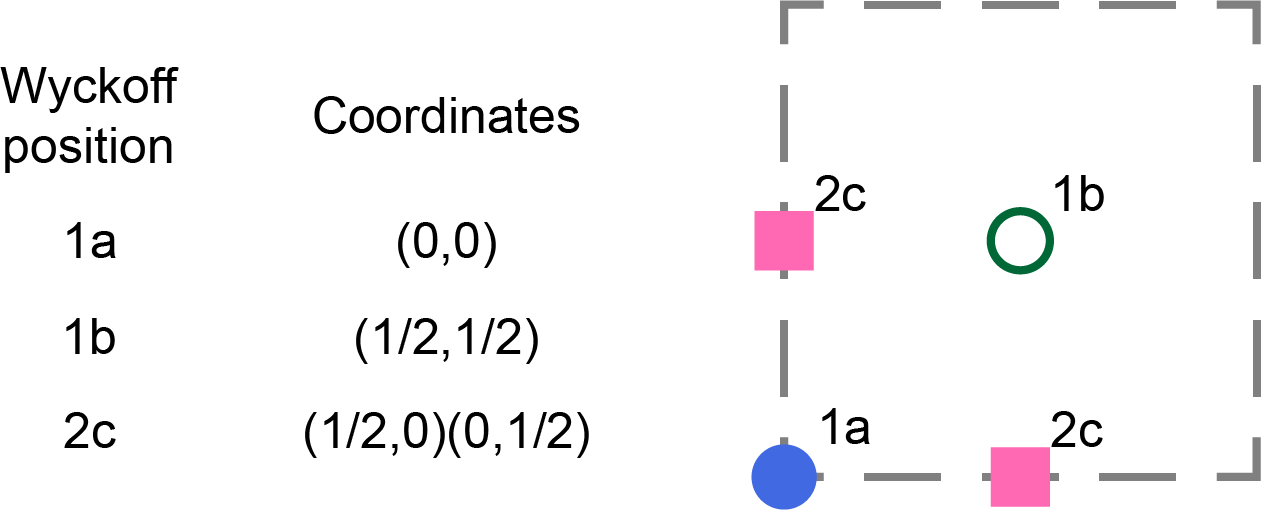}
    \caption{Maximal Wyckoff positions in the layer groups $p4/m1'$, $p4/m'$, $p4/m$, $p41'$, and $p4$. The general position $4d$ is not shown.
    }
    \label{fig_wyckoff}
\end{figure}

We now derive the filling anomaly in terms of the ion positions and Wannier centers of the occupied valence bands.
Both the ion positions and Wannier centers are described by Wyckoff positions; for a review of Wyckoff positions in the context of band theory, we refer the reader to Refs.~\cite{bradlyn2017topological,cano2018building}.
We use terminology specific to a finite square lattice of side length $L$, but the method is general.

The first step is to count the number of electrons in the charge neutral configuration.
To do this, we need to know the total number of ions in a finite-sized system with open boundary conditions.
Let $w$ be one of the four Wyckoff positions on the 2D square lattice, $1a$, $1b$, $2c$, or $4d$, shown in Fig.~\ref{fig_wyckoff}.
Define $N_w(L)$ to be the number of ions at the Wyckoff position $w$ that reside inside or on the boundary of a finite-sized square consisting of $L\times L$ unit cells.
For a periodic lattice of size $L\times L$, the number of ions is $L^2$ multiplied by the multiplicity of the Wyckoff position.
However, for open boundary conditions, this is not the case.
As shown in Fig.~\ref{fig_lattice}(b), for an $L\times L$ square with open boundary conditions:
\begin{align}
\label{eqn_N_w_ac}
    N_{a}(L)&=L^2 \nonumber\\
    N_b(L)&= (L-1)^2 \nonumber \\
    N_c(L) &= 2L(L-1).
\end{align}
The total number of electrons in the charge neutral configuration is then given by a sum over all Wyckoff positions:
\begin{equation}
\label{eqn_Nneu}
    N_{\rm neutral} = \sum_w N_w(L)a_w,
\end{equation}
where $a_w$ denotes the number of valence electrons from the ion at Wyckoff position $w$.
(More generally, if there are multiple ions not related by symmetry at the same Wyckoff position, which can happen for Wyckoff positions with a variable coordinate, such as the $4d$ position, then $a_w$ should be the sum of valence electrons from each symmetry-distinct ion in the Wyckoff position $w$.)

We now count the number of electrons required to symmetrically fill the Wannier centers of the valence bands in a finite-size system with open boundary conditions. 
Each Wannier center is labelled by a Wyckoff position, $w$, and an irreducible representation (irrep), $\rho_w$, of the site symmetry group of $w$.
(The site symmetry group of $w$ is the set of symmetry operations that leave $w$ invariant; therefore, Wannier functions centered at $w$ must transform as irreps of the site symmetry group. The irreps $\rho_w$ describe the symmetry of the Wannier functions.)
Let $n_{\rho_w}$ count the number of Wannier functions centered at $w$ that transform as $\rho_w$ and are not related by symmetry; in the language of band representations \cite{bradlyn2017topological,cano2018building}, $n_{\rho_w}$ counts the number of times the band representation labelled by $\rho_w$ appears in the valence bands.
Then the number of electrons needed to symmetrically fill the Wannier functions in a finite-size $L\times L$ square with open boundary conditions is:
\begin{equation}
\label{eqn_Nsym}
    N_{\rm symmetric} = \sum_{w} N_w(L) \sum_{\rho_w} n_{\rho_w} \text{dim}(\rho_w),
\end{equation}
where $\text{dim}(\rho_w)$ is the dimension of the irrep $\rho_w$. 

Eqs.~(\ref{eqn_Nneu}) and (\ref{eqn_Nsym}) define the filling anomaly:
\begin{align}
\label{eqn_eta_N_N}
    \eta &= N_{\rm neutral}-N_{\rm symmetric} \mod 4(\text{or}~8) \nonumber \\
         &= \sum_w N_w(L) (a_w-\sum_{\rho_w} n_{\rho_w}\text{dim}(\rho_w)) \mod 4(\text{or}~8),
\end{align}
where, as discussed in Sec.~\ref{sec_corresp}, mod~8 applies with time-reversal symmetry and mod~4 applies without.
Although this formula for $\eta$ includes an $L$-dependent term on the right hand side, the $L$-dependence disappears 
due to our assumptions that the system has no bulk charge polarization and no bulk net charge, which would contribute terms of order $L$ and $L^2$, respectively.
Thus, $\eta$ is independent of $L$.
A similar method to compute $\eta$ was used in Refs~\cite{schindler2019fractional,benalcazar2019quantization,watanabe2020corner}.

We now specify to the square lattice.
The general Wyckoff position, denoted $4d$, has coordinates $(x,y)$, where $x,y\neq 0,\frac{1}{2}$.
Ions on the $4d$ position always come in multiples of four, even on a finite size lattice, so that $N_{d}(L)$ must be a multiple of four.
Therefore, if time-reversal symmetry is absent, the term in Eq.~(\ref{eqn_eta_N_N}) coming from the $4d$ Wyckoff position is a multiple of four and does not contribute to the filling anomaly.
If time-reversal symmetry is present, all electrons come in Kramers pairs, causing the term in Eq.~(\ref{eqn_eta_N_N}) coming from the $4d$ Wyckoff position to be a multiple of eight, which again does not contribute to the filling anomaly.
Therefore, when computing the filling anomaly, we need only concern ourselves with the maximal Wyckoff positions $1a$, $1b$, and $2c$, shown in Fig.~\ref{fig_wyckoff}.

\begin{table}[b]
\begin{tabular}{ccc}
     Layer group & $w$ & Site-symmetry group\\
     \hline
     $p4/m1'$ & $1a, 1b$ & $4/m1'$ \\
            & $2c$ & $2/m1'$ \\
    \hline
    $p4/m'$ & $1a, 1b$ & $4/m'$ \\
            & $2c$ & $2/m'$ \\
    \hline
    $p4/m$ & $1a, 1b$ &  $4/m$\\
            & $2c$ & $2/m$ \\
    \hline
    $p4$ & $1a, 1b$ &  $4$\\
            & $2c$ & $2$\\
    \hline
    $p41'$ & $1a, 1b$ &  $41'$\\
            & $2c$ & $21'$
\end{tabular}
\caption{Site-symmetry groups for the maximal Wyckoff positions, $w$, of the layer groups $p4/m1'$, $p4/m'$, $p4/m$, $p4$, and $p41'$ in international notation \cite{litvin}. The site-symmetry groups for $1a$ and $1b$ are the same groups that leave $\Gamma$ and $M$ invariant in momentum space; similarly, the site-symmetry group for $2c$ leaves $X$ invariant.}
\label{tab_ssgroups}
\end{table}

\begin{table}[b]
\begin{tabular}{c|ccc}
$4/m1'$ &$E$&$C_4$&$I$\\
\hline
$E_{\frac12 g}$&$2$&$\sqrt{2}$&$2$\\
$E_{\frac12 u}$&$2$&$\sqrt{2}$&$-2$\\
$E_{\frac32 g}$&$2$&$-\sqrt{2}$&$2$\\
$E_{\frac32 u}$&$2$&$-\sqrt{2}$&$-2$\\
\end{tabular}
\quad
\begin{tabular}{c|cc}
$2/m1'$ &$E$&$I$\\
\hline
$E_{g}$&$2$&$2$\\
$E_{u}$&$2$&$-2$\\
\end{tabular}
\quad
\begin{tabular}{c|cc}
$4/m'$ or $41'$ &$E$&$C_4$\\
\hline
$E_{\frac12}$&$2$&$\sqrt{2}$\\
$E_{\frac32}$&$2$&$-\sqrt{2}$
\end{tabular}
\caption{Characters of the irreps of the point groups $4/m1'$, $2/m1'$, $4/m'$, and $41'$ with SOC.
Characters of $2/m'$ and $21'$ are not shown because they each have only one (two-dimensional) irrep with spin-orbit coupling.
The irreps are labelled in the notation of Ref.~\cite{altmann1994point}.
In all groups, $\cal T$ or $\cal{TI}$ requires all irreps consist of two-dimensional pairs with complex conjugate eigenvalues.
The characters of $C_2 \equiv C_4^2$ and $m_z \equiv C_4^2\cal{I}$ are always zero and not listed here.
The character tables are not square because we have not included all symmetry operations and, further, have not included the irreps without SOC.}
\label{tab_irrep}
\end{table}

\begin{table}[t]
\begin{tabular}{c|ccc}
$4$&$E$&$C_4$&$C_2$\\
\hline
${}^1E_{\frac12}$&$1$&$\epsilon$&$i$\\
${}^1E_{\frac32}$&$1$&$-\epsilon^*$&$-i$\\
${}^2E_{\frac32}$&$1$&$-\epsilon$&$i$\\
${}^2E_{\frac12}$&$1$&$\epsilon^*$&$-i$\\
\end{tabular}
\qquad
\begin{tabular}{c|cc}
$2$&$E$&$C_2$\\
\hline
${}^1E_{\frac12}$&$1$&$i$\\
${}^2E_{\frac12}$&$1$&$-i$\\
\end{tabular}
\caption{Characters of irreps of the point groups $4$ and $2$ with SOC; $\epsilon=\exp(\pi i/4)$.}
\label{tab_irrep_noT}
\end{table}

\begin{table}[t]
\begin{tabular}{c|cccc}
$4/m$ &$E$&$C_4$&$C_2$&$I$\\
\hline
${}^1E_{\frac12 g}$&$1$&$\epsilon$&$i$&$1$\\
${}^1E_{\frac32 g}$&$1$&$-\epsilon^*$&$-i$&$1$\\
${}^2E_{\frac32 g}$&$1$&$-\epsilon$&$i$&$1$\\
${}^2E_{\frac12 g}$&$1$&$\epsilon^*$&$-i$&$1$\\
${}^1E_{\frac12 u}$&$1$&$\epsilon$&$i$&$-1$\\
${}^1E_{\frac32 u}$&$1$&$-\epsilon^*$&$-i$&$-1$\\
${}^2E_{\frac32 u}$&$1$&$-\epsilon$&$i$&$-1$\\
${}^2E_{\frac12 u}$&$1$&$\epsilon^*$&$-i$&$-1$\\
\end{tabular}
\quad
\begin{tabular}{c|ccc}
$2/m$&$E$&$C_2$&$I$\\
\hline
${}^1E_{\frac12 g}$&$1$&$i$&$1$\\
${}^2E_{\frac12 g}$&$1$&$-i$&$1$\\
${}^1E_{\frac12 u}$&$1$&$i$&$-1$\\
${}^2E_{\frac12 u}$&$1$&$-i$&$-1$\\
\end{tabular}
\quad
\caption{Characters of irreps of $4/m$ and $2/m$ with SOC; $\epsilon=\exp(\pi i/4)$.}
\label{tab_irrep_I_noT}
\end{table}

A second simplification for the square lattice is that $\text{dim}(\rho_w) $ is independent of $\rho_w$ and $w$ for each symmetry group we consider, as we now explain.
The maximal Wyckoff positions and their site symmetry groups are listed in Table~\ref{tab_ssgroups} for all the layer groups we consider.
In $p4/m1'$, Table~\ref{tab_ssgroups} shows the only possible site-symmetry groups are $4/m1'$ or $2/m1'$; all irreps of these groups are two-dimensional, as shown in Table~\ref{tab_irrep}.
The same is true for $p4/m'$ and $p41'$.
In $p4$, Table~\ref{tab_ssgroups} shows the only possible site-symmetry groups are $4$ or $2$; all irreps of these groups are one-dimensional, as shown in Table~\ref{tab_irrep_noT}.
The same is true for $p4/m$; its irreps are enumerated in Table~\ref{tab_irrep_I_noT}.

Thus, in all cases, $\text{dim}(\rho_w)$ is independent of both $\rho$ and $w$ and the expression for $\eta$ in Eq.~(\ref{eqn_eta_N_N}) can be simplified on the square lattice as,
\begin{equation}
\label{eqn_eta_square}
    \eta \xrightarrow[\text{lattice}]{\text{square}} \sum_{w \text{~max}} N_w(L) (a_w-n_w d)~~ \text{mod}~4(\text{or}~8),
\end{equation}
where the sum is over all maximal Wyckoff positions $w$; $\eta$ is defined mod 4 (mod 8) in the absence (presence) of time-reversal symmetry, as in Eq.~(\ref{eqn_eta_N_N});
\begin{equation}
    n_w \equiv \sum_{\rho_w} n_{\rho_w};
    \label{eqn_def_nw}
\end{equation} 
and 
\begin{equation}
    \label{eqn_defd}
    d=\text{dim}(\rho_w) = \begin{cases} 2 & \text{ if }p4/m1', p4/m', p41'\\
                                            1 & \text{ if }p4/m, p4 \end{cases}
\end{equation} 
is the dimension of each irrep (which is independent of the choice of Wyckoff position and choice of irrep, as discussed in the previous paragraph.)

Plugging the formulas for $N_w(L)$ from Eq.~(\ref{eqn_N_w_ac}) into Eq.~(\ref{eqn_eta_square}) yields:
\begin{multline}
    \label{eqn_eta_in_nw}
    \eta \xrightarrow[\text{lattice}]{\text{square}} 
    L^2 \left[ N - d(n_a + n_b +2n_c) \right] \\ - 2L\left[ a_b+a_c - d(n_b+n_c)\right]  \\+ (a_b - dn_b) ~\text{mod}~4(\text{or}~8),
\end{multline}
where $\eta$ is defined mod 4 (mod 8) in the absence (presence) of time-reversal symmetry,
\begin{equation}
    N=a_a+a_b+2a_c 
    \label{eqn_defN}
\end{equation}
is the number of occupied bands, and $d$ is defined in Eq.~(\ref{eqn_defd}).

The bulk charge is determined by the number scaling with $L^2$ in Eq.~(\ref{eqn_eta_in_nw}) and must be zero in a system that is charge neutral in the bulk:
\begin{equation}
    N-d(n_a+n_b+2n_c) = 0 ~\text{mod}~4(\text{or}~8)
    \label{eqn_bulk0}
\end{equation}
The bulk polarization $p_x=p_y\equiv p$ is determined by the number scaling with $L$ in Eq.~(\ref{eqn_eta_in_nw}) and must also be zero:
\begin{equation}
    2\left[ a_b+a_c-d(n_b+n_c) \right] = 0~\text{mod}~4(\text{or}~8)
    \label{eqn_pol0}
\end{equation}
The filling anomaly $\eta$ is determined by the $L$-independent term in Eq.~(\ref{eqn_eta_in_nw}):
\begin{equation}
\label{eqn_eta_na}
    \eta = a_b-dn_b = a_a-dn_a ~\text{mod}~4(\text{or}~8),
\end{equation}
where the equality follows from Eqs.~(\ref{eqn_bulk0}) and (\ref{eqn_pol0}).
Equations (\ref{eqn_bulk0}), (\ref{eqn_pol0}) and (\ref{eqn_eta_na}) were also obtained in Ref.~\cite{watanabe2020corner}.

\subsection{\label{sec_2d_method_symm}Symmetry indicators for Wannier centers}

The formulas in the previous section derive the filling anomaly in terms of the crystal structure and Wannier centers.
We now derive $n_w$ in terms of the irreps of the little groups at high-symmetry points of the bulk band structure. This is useful because the irreps are easier to compute than the Wannier centers.
As we will see, because the irreps in momentum space do not completely determine the Wannier centers \cite{Bacry1988b,Michel1992,Bacry1993,bradlyn2017topological,cano2020band,Cano2020topology}, $n_w$ can only be determined up to some modulus from symmetry.


To this end, let $A$ be the integer ``EBR matrix'' of the symmetry group under consideration: each column of $A$ is labelled by an EBR and each row a particular irrep of the little group of a particular high-symmetry point.
The entries in the matrix indicate the number of times each momentum space irrep appears in the EBR \cite{song2020fragile,song2020twisted,cano2020band}.

A group of topologically trivial bands can be expressed as a linear combination of EBRs \cite{bradlyn2017topological} with integer coefficients $\tilde{n}_i$.
The irreps that appear at high-symmetry points in the band structure satisfy
\begin{equation}
    v=A\tilde{n},
    \label{eqn_vAn}
\end{equation}
where $v_j$ is the number of times the $j^\text{th}$ irrep appears in the band structure. 
We need to invert this equation to find $n_w$ in terms of $v$, as we now explain.

Let the Smith normal form of $A$ be given by
\begin{equation}
    A = U^{-1}DV^{-1},
    \label{eqn_smithA}
\end{equation}
where $D$ is a diagonal integer matrix with diagonal entries $(d_1, \dots, d_M, 0, \dots 0)$, i.e., the first $M$ entries are positive and the remaining entries are zero, and $U,V$ are integer matrices invertible over the integers.
(Note: the stable topological crystalline insulator classification of the group is given by $\otimes_{i=1}^M \mathbb{Z}_{d_i}$, where $\mathbb{Z}_{d_i}$ is the group of integers mod $d_i$ \cite{po2017symmetry,song2018quantitative,song2020fragile,song2020twisted,cano2020band}.)

We want to express the number of EBRs corresponding to each Wyckoff position in terms of symmetry irreps.
Since we are only considering topologically trivial bands, we consider only the vectors $v$ for which there exists an integer vector $\tilde{n}$ that solves Eq.~(\ref{eqn_vAn}).
Then the Smith normal form in Eq.~(\ref{eqn_smithA}) implies $Uv = (v_1',\dots v_M', 0 \dots 0)$, where $d_i$ divides $v_i'$.
For such bands, one solution to Eq.~(\ref{eqn_vAn}) is given by $\tilde{n}=VD^pUv$, where $D^p$ is the pseudoinverse of $D$, i.e., a diagonal matrix with diagonal entries $(1/d_1, 1/d_2, \dots, 1/d_l, 0, \dots 0)$.
This solution is not generically unique:
the most general solution to Eq.~(\ref{eqn_vAn}) is $\tilde{n}=VD^pUv + V\tilde{n}_0$, where $\tilde{n}_0$ is any integer vector in the nullspace of $D$, i.e., the first $M$ entries of $\tilde{n}_0$ are zero, so that $D\tilde{n}_0=0$.
Thus, given a particular $v$, $\tilde{n}_i$ can only be determined modulo $\text{gcd}\lbrace V_{ij} | _{j>M} \rbrace$, where gcd indicates the greatest common divisor.

However, we do not need each $\tilde{n}_i$ separately: we seek $n_w$ in Eq.~(\ref{eqn_def_nw}), which is a sum of all $\tilde{n}_i$ where the EBR indicated by $i$ is induced from an irrep of the site-symmetry group of the Wyckoff position $w$; we use $i\in w$ to denote this set of EBRs.
Then, following the previous paragraph, we can express $n_w$ as 
\begin{equation}
    n_w = \sum_{i\in w} \left[ VD^pUv\right]_i \mod \text{gcd}\lbrace \left( \sum_{i\in w} V_{ij} \right) | _{j > M} \rbrace.
    \label{eqn_smith_nw}
\end{equation}
We now use Eq.~(\ref{eqn_smith_nw}) to compute $n_w$ in $p4/m1'$ in terms of the symmetry indicators.
We do the same for $p4/m'$, $p4/m$, $p4$, and $p41'$ in Sec.~\ref{sec_2d_gen}.
The Smith normal form of the EBR matrix for each of these groups is computed in Appendix~\ref{app_smith}.

\subsubsection{Symmetry indicators for $n_{a,b,c}$ in $p4/m1'$}

In $p4/m1'$, there are three high-symmetry points in the Brillouin zone: $\Gamma= (0,0)$, $M=(\pi,\pi)$ and $X = (\pi,0)$. The point $(0,\pi)$ is symmetry-related to $X$, so does not add any new information. 
The points $\Gamma$ and $M$ are invariant under the point group $4/m1'$, while $X$ is invariant under $2/m1'$; the irreps of both groups are listed in Table~\ref{tab_irrep}.

Define $\# K_{\frac{1}{2}u}$ ($\# K_{\frac{3}{2}u}$) to be the number of times the irrep $E_{\frac{1}{2}u}$ ($E_{\frac{3}{2}u}$) appears in the valence band spectrum at the high-symmetry point $K= \Gamma, M$ and define $\# K_{u}=\# K_{\frac12 u}+\# K_{\frac32 u}$.
Similarly, define $\# X_u $ to be the number of times the irrep $E_u$ appears in the valence band spectrum at $X$.
Then define $[K_{\rho}]$ to be the difference between the number of times the irrep indicated by $\rho$ appears at the high-symmetry point $K$ and at $\Gamma$:
\begin{equation}
    \label{eqn_defKrho}
    [K_{\rho}]=\# K_{\rho}-\# \Gamma_{\rho}.
\end{equation}
Using this notation, we find from Eq.~(\ref{eqn_smith_nw}) (details in Appendix~\ref{app_smith}):
\begin{align}
\label{eqn_na}
    n_a &= \frac{N}{2}-[X_u]-\frac32[M_{u}]+2[M_{\frac12 u}] &\mod 4,\\
\label{eqn_nb}
    n_b &= [X_u]+\frac12 [M_{u}]-2[M_{\frac12 u}]&\mod 4,\\
\label{eqn_nc}
    n_c &= \frac12 [M_{u}] &\mod 2,
\end{align}
where $N=2(n_a+n_b+2n_c)$ is the total number of filled bands, which was derived by imposing bulk charge neutrality in Eq.~(\ref{eqn_bulk0}).

\subsection{Symmetry indicators for the filling anomaly}
\label{sec_2d_p4m1p}

We are now ready to compute the filling anomaly $\eta$ in Eq.~(\ref{eqn_eta_square}) in terms of the symmetry irreps by plugging in Eqs.~(\ref{eqn_na}), (\ref{eqn_nb}), and (\ref{eqn_nc}) for the group $p4/m1'$ (the results for other groups are in Sec.~\ref{sec_2d_gen}).
In previous work \cite{benalcazar2019quantization,schindler2019fractional}, $\eta$ was computed on 2D lattices with only one maximal Wyckoff position occupied.
The main advance of this work is to compute $\eta$ for square lattices with any number of atoms in the unit cell.
In Sec.~\ref{sec_2d_1}, we compute $\eta$ for the simple square lattice with one atom in the unit cell, reproducing earlier results \cite{schindler2019fractional}.
In Sec.~\ref{sec_2d_3}, we derive $\eta$ in the general case with multiple atoms in the unit cell.

\par

\subsubsection{Simple square lattices\label{sec_2d_1}}

We now reproduce the symmetry indicator formula in Ref.~\cite{schindler2019fractional} for a square lattice with one atom in the unit cell.
Without loss of generality, we can take that atom to be in the $1a$ position.
Then the formula for the filling anomaly in Eq.~(\ref{eqn_eta_na}) (with $d=2$) simplifies to
\begin{equation}
\label{eqn_eta_nb}
    \eta = -2n_b \mod 8.
\end{equation}
Plugging the expression for $n_b$ in Eq.~(\ref{eqn_nb}) into Eq.~(\ref{eqn_eta_nb}), we obtain the symmetry indicator formula:
\begin{align}
\label{eqn_eta_simple}
    \eta = -2[X_u]-[M_{u}]+4[M_{\frac12 u}] \mod 8.
\end{align}
Noticing that $[M_u]$ must be an even number in an (obstructed) atomic limit phase because $n_c$ in Eq.~(\ref{eqn_nc}) must be an integer, we deduce that $\eta$ is in fact a $\mathbb{Z}_4$ quantity. Eq.~(\ref{eqn_eta_simple}) was introduced in Eq.~(50) in Ref.~\cite{schindler2019fractional}.
\par

\subsubsection{General case: atoms in multiple Wyckoff positions\label{sec_2d_3}}

We now consider the general case, shown in Fig.~\ref{fig_lattice}(b), where there can be atoms at any Wyckoff positions.
Thus, the number of electrons from each ion, $a_{a,b,c}$, can all be non-zero. The number of filled bands is $N=a_a+a_b+2a_c$.
Plugging Eq.~(\ref{eqn_na}) into the expression for the filling anomaly in Eq.~(\ref{eqn_eta_na}) (with $d=2$) we find the symmetry indicator formula for the filling anomaly:
\begin{align}
\label{eqn_eta_mult}
    \eta &= a_a-N+2[X_u]+3[M_{u}]-4[M_{\frac12 u}] \mod 8.
\end{align}
Since time-reversal symmetry constrains $a_a$ and $N$ to be even numbers and, as discussed below Eq.~(\ref{eqn_eta_simple}), $[M_u]$ is also even, the filling anomaly is again a $\mathbb{Z}_4$ quantity.
When $N=a_a$, which implies $a_b = a_c = 0$, Eq.~(\ref{eqn_eta_mult}) is equivalent to Eq.~(\ref{eqn_eta_simple})
(to see this, notice that the equations are mod 8, $[M_u]$ is even, and when $a_b=a_c =0$, Eq.~(\ref{eqn_pol0}) implies $n_b = n_c \mod 2$, from which Eqs.~(\ref{eqn_nb}) and (\ref{eqn_nc}) together require that $[X_u]$ is also even.)
When $a_a\neq N$, Eq.~(\ref{eqn_eta_mult}) is distinct from Eq.~(\ref{eqn_eta_simple}) and has not appeared in previous literature.

In Sec.~\ref{sec_3d}, we build an explicit body-centered tetragonal model with $C_4$, $\cal{T}$ and $\cal{I}$ symmetry.
The Hamiltonian in the $k_z = 0$ and $k_z = \pi$ planes of the model describes a square lattice with $p4/m1'$ symmetry, but
with atoms at multiple Wyckoff positions, corresponding to the projection of the 3D model onto a 2D plane.
Therefore, the 2D bulk-corner correspondence derived in this section applies to 2D planes of that model, providing a numerical check of the analytical results.
\par

\subsection{\label{sec_2d_gen}Generalization to other layer groups}

We now compute the filling anomaly in terms of the symmetry indicators for the layer groups $p4/m'$, $p4$, and $p4/m$ and explain why $p41'$ does not have an analogous formula.
The formulas for $n_{a,b,c}$ in this section are derived in Appendix~\ref{app_smith}.

\subsubsection{$p4/m'$}
\label{sec_2d_p4mp}
For the layer group $p4/m'$, the high-symmetry points $\Gamma$ and $M$ are invariant under the point group $4/m'$, while $X$ is invariant under $2/m'$.
The irreps for these groups are listed in Table~\ref{tab_irrep}.
The number of Wannier centers at each Wyckoff position, $n_a$, $n_b$ and $n_c$, are defined by Eq.~(\ref{eqn_def_nw}) with irrep dimension $d=2$ in Eq.~(\ref{eqn_defd}). 
Using Eq.~(\ref{eqn_smith_nw}), we find $n_a$ and $n_b$ can be determined mod $2$:
\begin{align}
\label{eqn_na1}
    n_a &= \frac{N}{2}-[M_{\frac12 }]&\mod 2,\qquad\\
\label{eqn_nb1}
    n_b &= [M_{\frac12 }]&\mod 2,\qquad
\end{align}
where $[M_{\frac12}] = \# M_\frac12-\# \Gamma_\frac12$, where $\# K_\frac12$ indicates the number of times the irrep $E_\frac12$ appears in the valence bands at the high-symmetry point $K = \Gamma, M$. 
We find $n_c=0$ mod 1, i.e., $n_c$ is not constrained by symmetry irreps.

The filling anomaly is determined by Eq.~(\ref{eqn_eta_na}) taken mod 4 with $d=2$.
Substituting in Eq.~(\ref{eqn_na1}) yields:
\begin{equation}
\label{eqn_eta1}
    \eta_1 = a_a-N+2[M_{\frac12}]\mod 4,
\end{equation}
where the subscript is to distinguish the filling anomaly in $p4/m'$ from that computed in $p4/m1'$ in Eq.~(\ref{eqn_eta_mult}).
In the case where atoms only occupy one Wyckoff position ($1a$), $a_a=N$ and this equation reduces to
\begin{equation}
\label{eqn_eta11}
    \eta_1 = 2[M_{\frac12}]\mod 4.
\end{equation}
Eq.~(\ref{eqn_eta11}) was introduced in the context of higher order Fermi arcs \cite{wieder2020strong} with $C_4$ and $M_{x,y}$ symmetries: the anti-commuting reflection symmetries there play the same role as $\cal TI$ in $p4/m'$ in generating two dimensional irreps. \par

\subsubsection{$p4$}

For the layer group $p4$, the high-symmetry points $\Gamma$ and $M$ are invariant under the point group $4$, while $X$ is invariant under $2$; the irreps for these groups are listed in Table~\ref{tab_irrep_noT}.
The number of Wannier centers at each Wyckoff position, $n_a$, $n_b$ and $n_c$, are defined by Eq.~(\ref{eqn_def_nw}) with irrep dimension $d=1$ in Eq.~(\ref{eqn_defd}).
Using Eq.~(\ref{eqn_smith_nw}), we find that $n_a$ and $n_b$ can be determined mod $4$, while $n_c$ can be determined mod $2$: 
\begin{align}
    \label{eqn_na_2}
    n_a &= N-[X_2]+\frac32([M_1+M_3])+2[M_2] \mod 4\\
\label{eqn_nb_2}
    n_b &= [X_2]-\frac12 ([M_{1}]+[M_3])-2[M_2]~\mod 4\\
\label{eqn_nc_2}
    n_c &= -\frac12 ([M_1]+[M_3]) ~\mod 2
\end{align}
Here we use the notation $[M_j]=\# M_j-\# \Gamma_j$, where $j=1,2,3,4$ corresponds to the irrep with $C_4$ eigenvalue $\exp(i\frac{\pi}{2}(j-\frac12))$, and $[X_1]=\#X_1-\#\Gamma_1-\#\Gamma_3$, $[X_2]=\#X_2-\#\Gamma_2-\#\Gamma_4$, where $X_{1,2}$ corresponds to the irrep with $C_2$ eigenvalues $+i$, $-i$.
As in previous sections, $\# K_j$ indicates the number of times the irrep $j$ appears in the valence bands at the high-symmetry point $K$.

The expression for the filling anomaly is given by Eq.~(\ref{eqn_eta_na}) taken mod 4 with $d=1$:
\begin{align}
\label{eqn_eta_na2}
    \eta_2 &= a_a-n_a \mod 4 \\
         &= a_a-N+[X_2]-\frac32([M_1+M_3])-2[M_2] \mod 4,
         \label{eqn_eta2}
\end{align}
where Eq.~(\ref{eqn_eta2}) results from plugging Eq.~(\ref{eqn_na_2}) into Eq.~(\ref{eqn_eta_na2}).

\subsubsection{$p4/m$}

For the layer group $p4/m$, the high-symmetry points $\Gamma$ and $M$ are invariant under the point group $4/m$, while $X$ is invariant under $2/m$; the irreps for these groups are listed in Table~\ref{tab_irrep_I_noT}.
The number of Wannier centers at each Wyckoff position, $n_a$, $n_b$ and $n_c$, are defined by Eq.~(\ref{eqn_def_nw}) with irrep dimension $d=1$ in Eq.~(\ref{eqn_defd}).
From Eq.~(\ref{eqn_smith_nw}), we find that $n_a$ and $n_b$ can be determined mod $4$ while $n_c$ can be determined mod $2$. The formulas are:
\begin{align}
    \label{eqn_na_3}
    n_a &= N-[X_u]-\frac32[M_u]+2([M_{1u}]+[M_{2u}])\mod 4\\
\label{eqn_nb_3}
    n_b &= [X_u]+\frac12 [M_u]-2([M_{1u}]+[M_{2u}])~\mod 4\\
\label{eqn_nc_3}
    n_c &= \frac12 [M_u] ~\mod 2
\end{align}
Here we use the notation $[M_{j,\xi}]=\# M_{j,\xi}-\#\Gamma_{j,\xi}$, where the irrep of $4/m$ labelled by $j,\xi$ has 
$C_4$ eigenvalue $\exp(i\frac{\pi}{2}(j-\frac12))$, $j=1,2,3,4$,
and inversion eigenvalue $+1,-1$ corresponding to $\xi = g,u$.
In addition, $[X_{1,\xi}]=\# X_{1,\xi}-\#\Gamma_{1,\xi}-\#\Gamma_{3,\xi}$ and $[X_{2,\xi}]=\# X_{2,\xi}-\#\Gamma_{2,\xi}-\#\Gamma_{4,\xi}$, where $X_{1,\xi}$, $X_{2,\xi}$ correspond to the irreps of $2/m$ with $C_2$ eigenvalues $+i$, $-i$ and inversion eigenvalues $+1, -1$ corresponding to $\xi = g,u$. We denote $[K_\xi]=\sum_j[K_{j,\xi}]$. 

The filling anomaly is determined by Eq.~(\ref{eqn_eta_na}) taken mod 4 with $d=1$:
\begin{align}
\label{eqn_eta_na3}
    \eta_3 &= a_a-n_a \mod 4,\\
         &= a_a-N+[X_u]+\frac32[M_u]-2([M_{1u}]+[M_{2u}]) \mod 4,
         \label{eqn_eta3}
\end{align}
where Eq.~(\ref{eqn_eta3}) results from plugging Eq.~(\ref{eqn_na_3}) into Eq.~(\ref{eqn_eta_na3}).

\subsubsection{$p41'$}
\label{sec_p41p}

The group $p41'$ does not have a symmetry indicator formula, as we now explain.
The filling anomaly is given by Eq.~(\ref{eqn_eta_na}) taken mod 8 with $d=2$ (from Eq.~(\ref{eqn_defd})):
\begin{align}
    \label{eqn_eta_n4}
    \eta_4 &= a_a-2n_a \mod 8
\end{align}
However, $n_a$ is given by Eq.~(\ref{eqn_na1}) (as explained in Appendix~\ref{sec_Smith_p41p}, the symmetry indicator formula for $n_a$ is the same in $p41'$ as in $p4/m'$) and is only defined mod 2.
It follows from plugging Eq.~(\ref{eqn_na1}) into Eq.~(\ref{eqn_eta_n4}) that the symmetry indicator formula for $\eta_4$ is only defined mod 4, even though $\eta_4$ should be determined mod 8. 
Thus, we say that the symmetry indicator formula does not exist because the symmetry indicators do not provide enough information to completely determine the filling anomaly in this group.
The mod 8 filling anomaly $\eta_4$ can only be partially determined mod 4:
\begin{equation}
    \eta_4 \mod 4= a_a-N+2[M_{\frac12}]\mod 4.
    \label{eqn_eta4}
\end{equation}

\subsection{Summary of 2D results}
\label{sec_summary_2D}
Eqs.~(\ref{eqn_eta_mult}), (\ref{eqn_eta1}), (\ref{eqn_eta2}), and (\ref{eqn_eta3}) are the symmetry indicator formulas that express the filling anomaly in OALs in terms of the symmetry invariants. 
Together, these formulas encompass all square lattices with time-reversal, inversion symmetry, and/or their product,  
and any number of atoms in the unit cell.\par

All the formulas derived in this section are additive, and can be applied to an insulating band structure with any number of filled valence bands, as long as it is charge neutral, polarization free, and all the strong symmetry indicators are trivial. Because of the additivity, the formulas also apply to fragile topological phases, as discussed in Ref.~\cite{benalcazar2017quantized}.

We now discuss some connections to previous work.
As we have mentioned earlier, the formulas for the filling anomaly in Refs.~\cite{benalcazar2019quantization} and \cite{schindler2019fractional} do not accommodate multiple atoms in the unit cell.
In Ref.~\cite{watanabe2020corner}, the filling anomaly $\eta$ was computed in terms of $n_w$ in the general case of having multiple atoms in the unit cell, but $\eta$ was not expressed in terms of the symmetry indicators.
Finally, the real space invariants computed in Ref.~\cite{song2020twisted} are related to the $n_w$ computed here and are computed using the EBR matrix, but there is not a one-to-one correspondence between them.
Ref.~\cite{kooi2020bulk} also discusses real space topological invariants that go beyond symmetry indicators.

In Appendix~\ref{app_diamond}, we derive the symmetry indicator formula for the filling anomaly for a finite-sized square lattice with a boundary normal to the $(1,1)$ direction. It turns out that the formulas are the same as we have derived in this section, where the boundary is normal to the $(1,0)$ direction.
The results in this section can be generalized to a finite-sized square lattice with a boundary normal to any direction by defining a square supercell with a side parallel to the boundary.
Since the supercell necessarily contains multiple atoms, the results in Sec.~\ref{sec_2d_3} apply; in order to get the correct irreps at high-symmetry points, the band structure must be computed relative to the supercell.
Non-square terminations can also obey $C_4$ symmetry (for example, an octagon); while the general logic described in this section applies, the counting in Eq.~(\ref{eqn_N_w_ac}) will be different.

Interestingly, we have found numerically that the corner states survive even if the global $C_4$ symmetry is broken, i.e., in a square lattice with an integer number of unit cells.
Although rigorously the corner states are not protected if the lattice symmetry is broken globally, 
physically this makes sense because the localized states on one corner should not depend on how the lattice is terminated at other corners.
A different method to compute the presence of gapless boundary states in other systems with a fractional number of unit cells was discussed in Ref.~\cite{rhim2018unified}.

\begin{figure*}[t]
    \centering
    \includegraphics[width=0.9\linewidth]{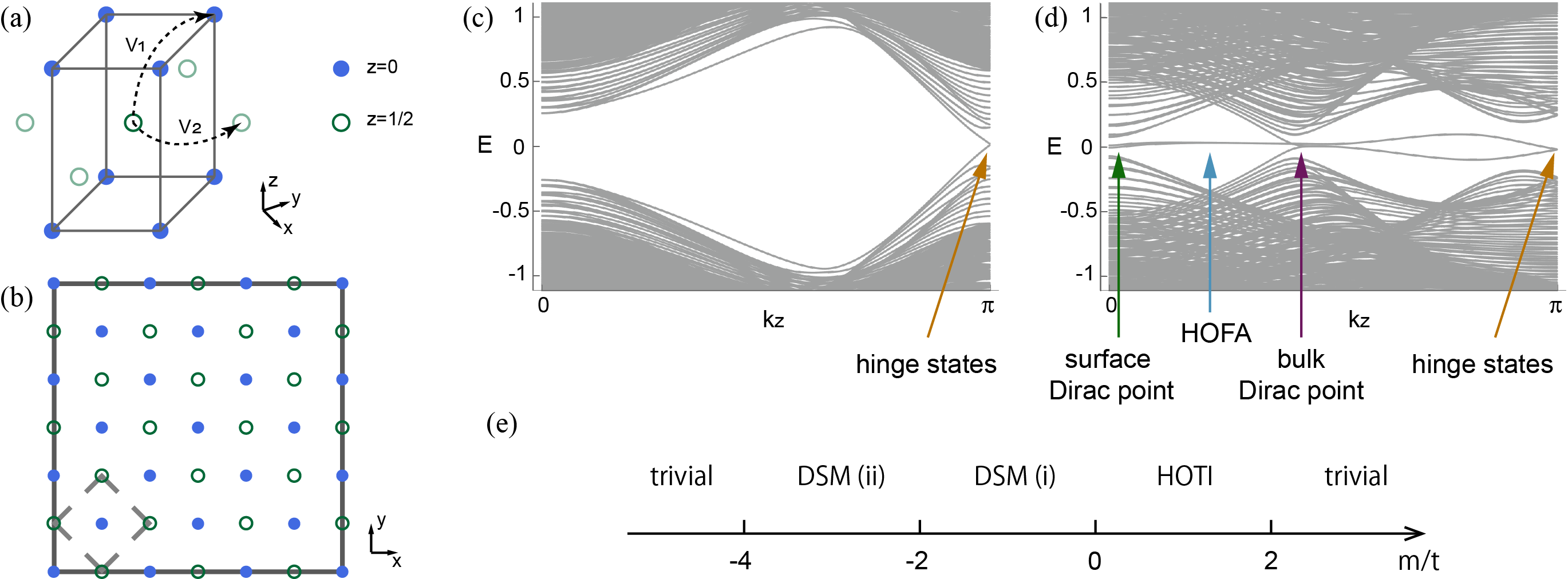}
    \caption{(a) Conventional unit cell of a body-centered tetragonal lattice;
    there are two sub-layers at $z=0,~\frac12$ (solid and hollow dots). 
    The eigenstates in the plane $(k_x,k_y,k_z=k_{z0})$ generically have weight on both sublattices.
    Therefore, the 2D Hamiltonian $H(k_x,k_y,k_{z0})$ describes atoms on a square lattice with two sites in the unit cell (b).
    (c) Spectrum for a rod (finite in the $x$- and $y$-directions, infinite in $z$) in the HOTI phase; gapless hinge states cross at $k_z=\pi$, forming an eight-fold degeneracy. (d) Spectrum for a rod in the DSM(i) phase, showing four-fold degenerate HOFA states between the surface Dirac point (projected at $k_z=0$) and bulk Dirac point. The HOFA regions have filling anomaly $\eta=2\mod 4$, while the crossing at $k_z=\pi$ has filling anomaly $\eta=4\mod 8$. (e) Phase diagram of our model, controlled by one parameter, $m/t$. }
    \label{fig_3d}
\end{figure*}

\section{3D model on the body-centered tetragonal lattice\label{sec_3d}}
We now apply the results derived in the previous section to classify the topology of a 3D tight-binding Hamiltonian.
We are interested in the HOTI and higher order Fermi arc (HOFA) phases. 
The topology of these phases can be understood by studying 2D slices of the Brillouin zone with fixed $k_z$. 
For example, the HOTI phase can be viewed as Wannier center pumping between the two time-reversal invariant planes ($k_z=0$ and $k_z=\pi$) \cite{song2017d}, while the HOFA phase requires each 2D fixed-$k_z$ slice on the Fermi arc to have a non-trivial filling anomaly \cite{wieder2020strong}. 
The difference between the previous works and the present manuscript is that we consider a $C_4$ symmetric body-centered tetragonal (BCT) model, whose 2D slices in momentum space 
necessarily contain atoms at different Wyckoff positions, as we will show in the next section.
Therefore, the results obtained in Sec.~\ref{sec_2d_3} and \ref{sec_2d_gen} are necessary to correctly identify the filling anomaly of 2D slices in momentum space.


\par

\subsection{Tight-binding model}
We build an explicit BCT tight-binding model in space group $87$ ($I4/m$) with a spin-$\frac32$ degree of freedom on each site.
The space group $I4/m$ is generated by body-centered lattice translations, a $C_4$ rotation about the $z$-axis, inversion, and time-reversal symmetry.
The matrix forms of these generators are given below in terms of the spin-$\frac32$ generators, $J_{x,y,z}$ (whose matrix forms are defined in Appendix~\ref{app_J}) and decomposed into Pauli matrices $\sigma_{0,x,y,z}$ and $\tau_{0,x,y,z}$:
\begin{align}
\label{eqn_C4}
    C_4 &= e^{-i\pi J_{z}/2}=\tau_z\sigma_z e^{-i\pi\sigma_z/4},\\
\label{eqn_Inv}
    {\cal I} &= \tau_0\sigma_0,\\
\label{eqn_TR}
    {\cal T} &= e^{-i\pi J_{y}}K=-i\tau_x\sigma_y K,
\end{align}
where $K$ is complex conjugation. 
It will be convenient to introduce $\tau_\pm = \frac12(\tau_x\pm i\tau_y)$. \par

We define a Cartesian coordinate system by the unit vectors $e_x=(1~0~0)a$, $e_y=(0~1~0)a$, $e_z=(0~0~1)c$, where $a$ and $c$ are lattice constants.
We align the base of the conventional tetragonal unit cell diagonally with respect to the $x$ and $y$ axes (see Fig.~\ref{fig_3d}(a)).
In this basis, the primitive translation vectors of the BCT lattice are given by
$e_1 = \frac12(e_y + e_z)$, $e_2 = \frac12(e_x+e_z)$, $e_3= \frac12(e_x + e_y)$.
The conventional unit cell has two atoms, indicated by solid and hollow circles in Fig.~\ref{fig_3d}(a), which form two sublattices. The primitive unit cell has one atom.

We take the lattice constants $c>a$. Therefore, each atom has four nearest neighbors on the same sublattice and eight next-nearest neighbors on the opposite sublattice. Our model only includes hopping between each site and these 12 nearest and next-nearest neighbors, illustrated in Fig.~\ref{fig_3d}(a). One of each of these hopping terms is given below (and drawn in Fig.~\ref{fig_3d}(a)) and the others are related by symmetry:
\begin{align}
    V_{0\rightarrow e_1}&=V_1=\frac14\left( \tau_z(t\sigma_z-\gamma\sigma_y)+\beta\tau_+\sigma_0+\beta^*\tau_-\sigma_0 \right) \nonumber \\
    V_{0\rightarrow e_3}&=V_2=-\frac14\left(2t\tau_z\sigma_z+\gamma\tau_y\sigma_0 \right) 
    \label{eqn_model}
\end{align}
The symmetry-related hopping terms in other directions are written explicitly in Appendix.~\ref{app_model}.
There is also an onsite mass term:
\begin{equation}
    V_{onsite} = m\tau_z\sigma_z 
    \label{eqn_model_mass}
\end{equation}
The parameters $m$, $t$, and $\gamma$ are real, and $\beta$ is complex; $\beta^*$ is the complex conjugate of $\beta$.
The Hamiltonian is written in momentum space in both the primitive and conventional bases in Appendix~\ref{app_model}. \par

\subsection{Topological phases}

We now classify the topology of this model by considering the Hamiltonian in 2D slices of the 3D Brillouin zone. 
Specifically, for fixed $k_{z0}$, $H(k_x,k_y,k_{z0})$ can be regarded as the Hamiltonian of a 2D system. 
Since eigenstates generically have weight on both sublattices, this 2D Hamiltonian describes atoms on a square lattice with two sites in the unit cell, as shown in Fig.~\ref{fig_3d}(b). If the 2D Hamiltonian has no gapless edge states or polarization, then we can apply our results from Sec.~\ref{sec_2d_3} and \ref{sec_2d_gen} to determine the corner charge of this 2D model when the 3D system is truncated in the $x$ and $y$ directions but infinite in the $z$ direction.

We find that our model has several topological phases depending on the ratio $m/t$. The parameters $\beta$ and $\gamma$ do not change the topological phase. We list all the different phases in Fig.~\ref{fig_3d}(e) with respect to $m/t$: there is one HOTI phase and two Dirac semimetal (DSM) phases, which we now describe.

\subsubsection{$\mathbb{Z}_8$ HOTI}

\begin{table}[]
    \centering
    \begin{tabular}{|c|c|c|c|c|c|c|c|c|c|}
    \hline
         $k_z$& $a_a$ & $N$ & $[X_u]$ & $[M_u]$ & $[M_{\frac12 u}]$ &$n_a$ &$n_b$ & $n_c$ &$\eta$\\
         \hline
         0& 2 & 4 & 1 & 0 & 0 & 1 & 1 & 0 &0\\
         \hline
         $\pi$& 2 & 4 & -1 & 0 & 0 & 3 & -1 & 0 & 4\\
    \hline
    \end{tabular}
    \caption{Symmetry indicators of occupied bands, $n_{a,b,c}$, and $\eta$ in the $k_z=0$ and $k_z=\pi$ planes of our BCT model in the HOTI phase. The filling anomalies $\eta$ are computed by plugging the values in the table into Eq.~(\ref{eqn_eta_mult}), while $n_{a,b,c}$ are computed from Eqs.~(\ref{eqn_na}), (\ref{eqn_nb}) and (\ref{eqn_nc}). The irreps are derived in Appendix~\ref{sec_3d_irreps}, Table~\ref{tab_calculation_HOTI}. 
    Notice that $[X_u]$ need not be even because although the $(1,1)$ boundary shares the same  symmetry indicator formula as the $(1,0)$ boundary, the constraints from zero bulk charge polarization and zero bulk net charge are different, as derived in Appendix~\ref{app_diamond}.
    }
    \label{tab_BCT}
\end{table}

When $0< m/t < 2$, a $\mathbb{Z}_8$ HOTI with the non-trivial symmetry indicator $\Delta=4$ is realized.
This phase was introduced in Refs.~\cite{song2017d,khalaf2018symmetry}.
While all the $k_z$-slices are either atomic limits or fragile phases in 2D, 
the 3D phase is stable topological because the 2D Wannier centers move as a function of $k_z$, which leads to helical modes on 1D hinges where the $x$- and $y$-normal surfaces meet.

Here, we derive this 3D phase on the BCT lattice by computing the filling anomaly in the $k_z=0$ and $k_z=\pi$ planes.
Since these two planes are invariant under time reversal, they are described by the layer group $p4/m1'$, for which we derived the symmetry indicator formula for the filling anomaly in Eq.~(\ref{eqn_eta_mult}) of Sec.~\ref{sec_2d_p4m1p}.

The symmetry indicators are shown in Table~\ref{tab_BCT}, from which the filling anomaly can be computed with Eq.~(\ref{eqn_eta_mult}).
We find that $\eta=4\mod 8$ in the $k_z = \pi$ plane and $\eta = 0$ in the $k_z=0$ plane.
The non-trivial value in the $k_z=\pi$ plane is responsible for the eight-fold degenerate mid-gap states we observe in Fig.~\ref{fig_3d}(c) in the $k_z=\pi$ plane for a finite-size rod geometry, which is finite in the $x$- and $y$-directions and infinite in the $z$ direction.
Since $\eta=0$ in the $k_z=0$ plane, to continuously connect the mid-gap states with the rest of the band structure requires $k_z$-dependent modes that traverse the bulk band gap; these are exactly the helical hinge modes required by the 3D HOTI phase.
Interestingly, Table~\ref{tab_BCT} shows that $n_b=-1$; thus, this slice is a fragile topological 2D insulator.
(This by itself is not enough to show that the slice is fragile because $n_{a,b}$ are only defined mod 4 (and $n_c$ mod 2); some algebra shows that there is no solution where $n_{a,b,c}>0$.)

We now reiterate the importance of our analysis in correctly describing this phase: our formula~(\ref{eqn_eta_mult}) correctly captures the filling anomaly in the $k_z=\pi$ plane (and lack of filling anomaly in the $k_z = 0$ plane), which agrees with both our numerical results and the helical edge modes predicted by the 3D formula in Ref.~\cite{song2017d}.
Earlier formulas for the filling anomaly give an incorrect result.
Specifically, Eq.~(\ref{eqn_eta_simple}) of Ref.~\cite{schindler2019fractional} yields {$\eta=6$} in the $k_z=0$ plane and {$\eta=2$} in the $k_z = \pi$ plane. 
This discrepancy occurs because Ref.~\cite{schindler2019fractional} does not accommodate atoms at multiple Wyckoff positions. 
While $\eta = 2$ and $\eta = 6$ are consistent with a helical mode, they do not agree with the state counting in our numerics, where there is always the same number of occupied and empty bands (and hence only consistent with $\eta = 0$ and $\eta = 4$, as we correctly obtain from Eq.~(\ref{eqn_eta_mult})).

\subsubsection{HOFA in Dirac Semimetals}

\begin{table}[]
    \centering
    \begin{tabular}{|c|c|c|c|c|c|c|c|}
    \hline
         $k_z$& $a_a$ & $N$ & $[M_{\frac12 }]$ &$n_a$ &$n_b$ & $n_c$ &$\eta$\\
         \hline
         $ 0<k_z< k_0 $& 2 & 4 & 0  & 2 & 0 & 0 &2\\
         \hline
         $k_0 < k_z < \pi $& 2 & 4 & 1 & 1 & 1 & 0 & 0\\
    \hline
    \end{tabular}
    \caption{In the two DSM phases, the bulk Dirac point is projected to $k_z=k_0$. For the two regions $ 0<k_z< k_0 $ and $k_0 < k_z < \pi$, the filling anomalies $\eta$ are computed by plugging the values in the table into Eq.~(\ref{eqn_eta1}), while $n_{a,b}$ are computed from Eqs.~(\ref{eqn_na1}) and (\ref{eqn_nb1}) ($n_c$ is always zero mod 1, as discussed below Eq.~(\ref{eqn_nb1})).
    The symmetry indicators are derived from the irreps listed in Table~\ref{tab_calculation_DSMi} of Appendix~\ref{sec_3d_irreps}.
    }
    \label{tab_BCT_HOFA}
\end{table}

When $-4 < m/t < 0$, the model is in one of two DSM phases and has two Dirac points along $\Gamma-Z$ in the bulk. 
The band structure for the DSM(i) phase on a rod finite in the $x$- and $y$-directions and infinite in $z$ is shown in Fig.~\ref{fig_3d}(d).
For both of the phases, there are HOFA hinge states connecting the projection of the two bulk Dirac points. The HOFA hinge states pass through $k_z=0$, where there are also projected gapless mirror Chern surface states. The difference between the two phases is that the phase DSM(i) has filling anomaly $\eta = 4$ mod 8 at $k_z=\pi$ while the plane $k_z=\pi$ in phase DSM(ii) is trivial.
\par

The HOFA hinge states occur in the planes between the two Dirac points.
Since these planes are not time-reversal invariant, but are invariant under the product of time-reversal and inversion, they are described by the layer group $p4/m'$, for which we derived the symmetry indicator formula for the filling anomaly in Eq.~(\ref{eqn_eta1}) of Sec.~\ref{sec_2d_p4mp}.
Applying this formula, we find $\eta=2$ in the planes between the bulk Dirac points, as derived in Table~\ref{tab_BCT_HOFA}.

The HOFA are not correctly described by formulas in previous work derived by assuming atoms at only one maximal Wyckoff position: for example, Eq.~(\ref{eqn_eta11}) yields $\eta = 0$ 
in these planes, which would indicate a lack of hinge modes.

In the DSM(i) phase, besides the HOFA, there are two groups of corner-localized hinge states that cross at $k_z=\pi$.  These states can be pushed into the valence or conduction bands by adding some symmetry protecting potentials, however the filling anomaly at $k_z=\pi$ remains non-trivial. These states are not present in the DSM(ii) phase which has $\eta = 0$ in the $k_z = \pi$ plane.

Further, we note that the gapless surface states at $k_z=0$ are unavoidable, even if the protecting mirror symmetry is broken, because the HOFA states are four fold degenerate while the only possible degeneracy of mid gap states at a time-reversal symmetric plane is eight (corresponding to a Kramers pair of time-reversed partners at each corner.)
This discrepancy between the HOFA degeneracy and the required degeneracy at $k_z = 0$ can only be resolved by having gapless surface or bulk states projected to the point. 

\section{discussion\label{sec_discussion}}

In this manuscript, we introduced a general method to derive the symmetry indicator formula for the filling anomaly that applies to crystals with any number of atoms in the unit cell.
We introduced an algorithm using the Smith normal form that makes the derivation systematic.
We applied this method to derive the filling anomaly on the 2D square lattice with time-reversal, inversion, and/or their product.
We further showed where our results go beyond earlier work that did not apply to crystals with atoms occupying multiple maximal Wyckoff positions.


We verified our results by correctly predicting the helical hinge modes and HOFAs in a 3D BCT tight-binding model, by analyzing 2D slices of the Brillouin zone.
This model served as a concrete example where previous results break down, showing the importance of our extension to crystals with atoms in multiple maximal Wyckoff positions.

Our method can be easily generalized to other $C_n$ ($n=2,3,4,6$) symmetric 2D lattices and higher dimensional lattices. 
It will also be interesting to apply our results to other topological semimetals, such as those in \cite{bradlyn2016beyond} which occur on non-symmorphic lattices and therefore necessarily have multiple atoms in the unit cell.\par

\textit{Note added.} During the final stages of this work, Ref.~\cite{takahashi2021general} appeared on the ArXiv, which also computes the filling anomaly in terms of symmetry indicators for general lattices with $C_n$ symmetry. 
Our results agree where they overlap.

\begin{acknowledgments}

This material is based upon work supported by the National Science Foundation under Grant No. DMR-1942447.
J.C. acknowledges the support of the Flatiron Institute, a division of the Simons Foundation.
\end{acknowledgments}

\bibliography{reference.bib}

\begin{appendix}

\section{\label{app_smith} Symmetry indicator formulas for the number of occupied EBRs at each Wyckoff position}

In this Appendix, we explain how the formulas for $n_{a,b,c}$ in the main text are derived from the Smith normal form.

\subsection{$p4/m1'$}
\begin{table}[b]
    \centering
    \begin{tabular}{|c|c|c|c|c|}
         \hline
         $w$ & $\rho$ &  $\Gamma$ & $X$ & $M$\\
         \hline
        $1a$ & $(p,\xi)$ & $(p,\xi)$ &$(\xi)$ & $(p,\xi)$\\
        \hline
        $1b$ & $(p,\xi)$ & $(p,\xi)$ &$(-\xi)$ & $(-p,\xi)$\\    
        \hline
        $2c$ & $\xi$ & $(p,\xi),(-p,\xi)$ &$(\xi),(-\xi)$ & $(p,-\xi),(-p,-\xi)$\\
        \hline
    \end{tabular}
    \caption{EBRs in $p4/m1'$. The Wyckoff positions, $w$, are listed in the first column.
    Each EBR is induced from an irrep, $\rho$, of the site-symmetry group of a Wyckoff position: as shown in Table~\ref{tab_ssgroups}, the site-symmetry group of the $1a$ and $1b$ positions is $4/m1'$, while the site-symmetry group of the $2c$ position is $2/m1'$.
    Irreps of $4/m1'$ can be labelled by a pair of $C_4$ eigenvalues, $\exp(\pm i p\pi/2)$, where $p=\frac12$ or $\frac32$, and their inversion eigenvalue, $\xi$; the pairs of $(p,\xi)$ correspond to the subscripts $\frac12 g, \frac12 u, \frac32 g, \frac32 u$ in Table~\ref{tab_irrep}.
    Irreps of $2/m1'$ can be distinguished by only their inversion eigenvalue, $\xi$, corresponding to the subscript $g$ or $u$ in Table~\ref{tab_irrep}.
    The labels $p$ (where applicable) and $\xi$ are indicated in the second column.
    The remaining columns indicate the irreps of the EBR in momentum space: $\Gamma$ and $M$ are invariant under $4/m1'$, while $X$ is invariant under $2/m1'$.   
    }
    \label{tab_EBR}
\end{table}


We first consider the group $p4/m1'$. The band representations induced from the three maximal Wyckoff positions $1a$, $1b$, and $2c$ (shown in Fig.~\ref{fig_wyckoff}) are listed in Table~\ref{tab_EBR}. 
Each band representation is expressed as a vector $v$ in the basis:
\begin{equation}
\label{eqn_basis_br}
    \left( E^{\Gamma}_{\frac12g}, E^{\Gamma}_{\frac12u}, E^{\Gamma}_{\frac32g}, E^{\Gamma}_{\frac32u}, E^{X}_{g}, E^{X}_{u}, E^{M}_{\frac12g}, E^{M}_{\frac12u}, E^{M}_{\frac32g}, E^{M}_{\frac32u} \right) ,
\end{equation}
where $E_\rho^K$ indicates the number of times the irrep $\rho$ occurs at the high-symmetry point $K$ in the momentum-space band representation. Note $\Gamma$ and $M$ are invariant under $4/m1'$, while $X$ is invariant under $2/m1'$; their irreps are defined in Table~\ref{tab_irrep}. 

Each group of topologically trivial bands, isolated in energy from all other bands, can be written as an integer linear combination of EBRs.
The coefficients form a vector $n$ in the following basis:
\begin{equation}
\label{eqn_basis_ebr}
    \left( E^{1a}_{\frac12g}, E^{1a}_{\frac12u}, E^{1a}_{\frac32g}, E^{1a}_{\frac32u}, E^{1b}_{\frac12g}, E^{1b}_{\frac12u}, E^{1b}_{\frac32g}, E^{1b}_{\frac32u}, E^{2c}_{g}, E^{2c}_{u} \right) ,
\end{equation}
where $E^{w}_{\rho}$ indicates the number of times the EBR induced from the two dimensional irrep $\rho$ of the site-symmetry group of the Wyckoff position $w$ appears in the linear combination.
In this basis, we use Table~\ref{tab_EBR}, which lists all EBRs and their momentum space irreps, to construct the EBR matrix defined in Sec.~\ref{sec_2d_method_symm}:
\begin{equation}
    \label{eqn_EBRmat}
    A = \begin{pmatrix}
    1 & 0 & 0 & 0 & 1 & 0 & 0 & 0 & 1 & 0 \\
 0 & 1 & 0 & 0 & 0 & 1 & 0 & 0 & 0 & 1 \\
 0 & 0 & 1 & 0 & 0 & 0 & 1 & 0 & 1 & 0 \\
 0 & 0 & 0 & 1 & 0 & 0 & 0 & 1 & 0 & 1 \\
 1 & 0 & 1 & 0 & 0 & 1 & 0 & 1 & 1 & 1 \\
 0 & 1 & 0 & 1 & 1 & 0 & 1 & 0 & 1 & 1 \\
 1 & 0 & 0 & 0 & 0 & 0 & 1 & 0 & 0 & 1 \\
 0 & 1 & 0 & 0 & 0 & 0 & 0 & 1 & 1 & 0 \\
 0 & 0 & 1 & 0 & 1 & 0 & 0 & 0 & 0 & 1 \\
 0 & 0 & 0 & 1 & 0 & 1 & 0 & 0 & 1 & 0 \\
    \end{pmatrix}
\end{equation}
Each column of $A$ indicates the momentum space irreps of a particular EBR, where the columns are ordered according to the list of EBRs in (\ref{eqn_basis_ebr}) and the rows are ordered according to the list of momentum space irreps in (\ref{eqn_basis_br}).
For example, the first column corresponds to the band representation induced by $E^{1a}_{\frac12 g}$ and the first entry, ``1'', indicates that the irrep $E^{\Gamma}_{\frac12g}$ appears one time in this EBR (as listed in Table~\ref{tab_EBR}).

Following Sec.~\ref{sec_2d_method_symm}, we apply the Smith decomposition to the EBR matrix $A$:
\begin{equation}
    \label{eqn_smith}
    A = U^{-1}DV^{-1}.
\end{equation}
where $U$ and $V$, which are invertible over integers, are found to be:
\begin{equation}
    \label{eqn_smith_U}
   U = \begin{pmatrix}
    0 & 0 & 0 & 0 & 0 & 0 & 1 & 0 & 0 & 0 \\
 0 & 0 & 0 & 0 & 0 & 0 & 0 & 1 & 0 & 0 \\
 0 & 0 & 1 & 0 & 0 & 0 & 0 & 0 & 0 & 0 \\
 0 & 0 & 0 & 1 & 0 & 0 & 0 & 0 & 0 & 0 \\
 0 & 0 & 0 & -1 & 0 & 1 & 0 & -1 & 0 & 0 \\
 1 & 1 & 1 & 0 & 0 & 0 & -1 & -1 & -1 & 0 \\
 0 & 0 & 1 & -1 & 0 & 1 & 0 & -1 & -1 & 0 \\
 1 & 0 & -1 & 2 & 0 & -2 & -1 & 2 & 1 & 0 \\
 -1 & -1 & -1 & -1 & 1 & 1 & 0 & 0 & 0 & 0 \\
 -1 & -1 & -1 & -1 & 0 & 0 & 1 & 1 & 1 & 1 \\
    \end{pmatrix}
\end{equation}
\begin{equation}
    \label{eqn_smith_V}
   V = \begin{pmatrix}
    1 & 0 & 0 & 0 & 0 & 0 & 0 & 1 & -1 & -1 \\
 0 & 1 & 0 & 0 & 0 & 0 & -1 & -2 & -1 & -1 \\
 0 & 0 & 1 & 0 & 0 & 0 & -1 & -1 & -1 & -1 \\
 0 & 0 & 0 & 1 & 0 & 0 & 0 & 0 & -1 & -1 \\
 0 & 0 & 0 & 0 & 1 & 0 & 0 & 1 & 1 & 0 \\
 0 & 0 & 0 & 0 & 0 & 1 & -1 & -2 & 1 & 0 \\
 0 & 0 & 0 & 0 & 0 & 0 & 0 & -1 & 1 & 0 \\
 0 & 0 & 0 & 0 & 0 & 0 & 0 & 0 & 1 & 0 \\
 0 & 0 & 0 & 0 & 0 & 0 & 1 & 2 & 0 & 1 \\
 0 & 0 & 0 & 0 & 0 & 0 & 0 & 0 & 0 & 1 \\
    \end{pmatrix}
\end{equation}
and the diagonal matrix $D$ is
\begin{equation}
\label{eqn_smith_diag}
    D=\text{diag}(1,1,1,1,1,1,1,4,0,0).
\end{equation}
The eighth entry, 4, which is the only non-zero, non-unity entry of $D$, indicates the $\mathbb{Z}_4$ symmetry indicator classification of this group \cite{po2017symmetry}.
The zero entries impose the constraint that an insulator must have the same number of occupied bands at all high-symmetry points.

To see this last point, consider a vector $v$ in the basis of Eq.~(\ref{eqn_basis_br}), which satisfies $v=A\tilde{n}$ for some integer vector $\tilde{n}$ that represents a sum of EBRs in the basis of Eq.~(\ref{eqn_basis_ebr}). According to the Smith decomposition in Eq.~(\ref{eqn_smith}), $Uv=DV^{-1}\tilde{n}$. Since the ninth and tenth entries on the diagonal of $D$ are zero, it follows that $\left[Uv\right]_{9,10} = 0$.
Plugging in the entries of $U$ from Eq.~(\ref{eqn_smith_U}) and using the basis of $v$ in Eq.~(\ref{eqn_basis_br}) yields two equations:
\begin{align}
   0 & = -E^{\Gamma}_{\frac12g}- E^{\Gamma}_{\frac12u}-E^{\Gamma}_{\frac32g}-E^{\Gamma}_{\frac32u}+E^{X}_{g}+E^{X}_{u} \nonumber\\
   0 &= -E^{\Gamma}_{\frac12g}- E^{\Gamma}_{\frac12u}- E^{\Gamma}_{\frac32g}- E^{\Gamma}_{\frac32u}+E^{M}_{\frac12g}+ E^{M}_{\frac12u}+ E^{M}_{\frac32g}+ E^{M}_{\frac32u}.
\end{align}
The first line specifies that there must be the same number of occupied bands at $\Gamma$ as at $X$ and the second line specifies that there must be the same number of occupied bands at $\Gamma$ and at $M$.

The pseudoinverse of $D$ is
\begin{equation}
    D^p=\text{diag}(1,1,1,1,1,1,1,1/4,0,0).
\end{equation}
We can now plug $U$, $V$, and $D^p$ into Eq.~(\ref{eqn_smith_nw}) to find $n_{a,b,c}$ in Eqs.~(\ref{eqn_na}), (\ref{eqn_nb}) and (\ref{eqn_nc}).

Physically, the ambiguity in the modulus of $n_{a,b,c}$ comes from the fact that the Wannier centers are not gauge invariant; this point is elaborated on in Ref.~\cite{song2020twisted}.
For example, one can check (by using $v=A\tilde{n}$) that the irreps at high-symmetry points for the band representations represented by $\tilde{n}=(1111000000)$, $\tilde{n}=(0000111100)$ and $\tilde{n}=(0000000011)$ are identical. This corresponds to the fact that the Wannier centers for these three band representations can all be continuously moved to the general Wyckoff position $4d$ without breaking symmetries. Therefore, they are physically indistinguishable by symmetry indicators.

\subsection{$p4/m'$}
\label{sec_Smith_p4mp}

The group $p4/m'$ contains $C_4$ and $\cal TI$ symmetries, but not $\cal T$ or $\cal I$ separately.
A basis for its irreps and EBRs can be read from the previous subsection by forgetting about inversion symmetry. 
Specifically, the basis for the irrep labels in momentum space is:
\begin{equation}
    \left( E^{\Gamma}_{\frac12},E^{\Gamma}_{\frac32},E^{X}_{\frac12},E^{M}_{\frac12},E^{M}_{\frac32} \right),
\end{equation}
which is the same as Eq.~(\ref{eqn_basis_br}) without the $g,u$ labels for the inversion eigenvalue, and
the basis for the multiplicity of each EBR is:
\begin{equation}
    \left( E^{1a}_{\frac12},E^{1a}_{\frac32},E^{1b}_{\frac12},E^{1b}_{\frac32},E^{2c}_{\frac12} \right),
\end{equation}
which is the same as Eq.~(\ref{eqn_basis_ebr}) without the $g, u$ labels.
The EBR matrix is:
\begin{equation}
    A = \begin{pmatrix}
    1 & 0 & 1 & 0 & 1 \\
 0 & 1 & 0 & 1 & 1 \\
 1 & 1 & 1 & 1 & 2 \\
 1 & 0 & 0 & 1 & 1 \\
 0 & 1 & 1 & 0 & 1 \\
    \end{pmatrix}
\end{equation}
The Smith decomposition (Eq.~(\ref{eqn_smith})) yields the matrices:
\begin{equation}
    D = \text{diag}(1,1,1,0,0),
\end{equation} 
\begin{equation}
   U = \begin{pmatrix}
    0 & 0 & 0 & 1 & 0 \\
 0 & 1 & 0 & 0 & 0 \\
 1 & 0 & 0 & -1 & 0 \\
 -1 & -1 & 1 & 0 & 0 \\
 -1 & -1 & 0 & 1 & 1 \\
    \end{pmatrix},
\end{equation}
\begin{equation}
   V = \begin{pmatrix}
    1 & 0 & 0 & -1 & -1 \\
 0 & 1 & 0 & -1 & -1 \\
 0 & 0 & 1 & 1 & 0 \\
 0 & 0 & 0 & 1 & 0 \\
 0 & 0 & 0 & 0 & 1 \\
    \end{pmatrix}
\end{equation}

We then apply Eq.~(\ref{eqn_smith_nw}) to derive Eqs.~(\ref{eqn_na1}) and (\ref{eqn_nb1}) for $n_a$ and $n_b$. 
Applying Eq.~(\ref{eqn_smith_nw}) to $n_c$ shows that it is only determined mod 1,  which does not provide any new information.

\subsection{$p4$}

In the case of $p4$, the site-symmetry groups of the maximal Wyckoff positions are defined in Table~\ref{tab_ssgroups} and their irreps are defined in Table~\ref{tab_irrep_noT}. The basis of irreps in momentum space is:
\begin{equation}
    \left(
    {}^1E_{\frac12}^{\Gamma}, {}^1E_{\frac32}^{\Gamma}, {}^2E_{\frac12}^{\Gamma}, {}^2E_{\frac32}^{\Gamma}, {}^1E_{\frac12}^{X}, {}^2E_{\frac12}^{X}, {}^1E_{\frac12}^{M}, {}^1E_{\frac32}^{M}, {}^2E_{\frac12}^{M}, {}^2E_{\frac32}^{M}
    \right)
\end{equation}
The basis of EBRs is:
\begin{equation}
\left(
    {}^1E_{\frac12}^{1a}, {}^1E_{\frac32}^{1a}, {}^2E_{\frac12}^{1a}, {}^2E_{\frac32}^{1a}, {}^1E_{\frac12}^{1b}, {}^1E_{\frac32}^{1b}, {}^2E_{\frac12}^{1b}, {}^2E_{\frac32}^{1b}, {}^1E_{\frac12}^{2c}, {}^2E_{\frac12}^{2c}
\right)
\end{equation}
Then the EBR matrix is
\begin{equation}
    \label{eqn_EBRmat_3}
    A = \begin{pmatrix}
    1 & 0 & 0 & 0 & 1 & 0 & 0 & 0 & 1 & 0 \\
 0 & 1 & 0 & 0 & 0 & 1 & 0 & 0 & 0 & 1 \\
 0 & 0 & 1 & 0 & 0 & 0 & 1 & 0 & 1 & 0 \\
 0 & 0 & 0 & 1 & 0 & 0 & 0 & 1 & 0 & 1 \\
 1 & 0 & 1 & 0 & 0 & 1 & 0 & 1 & 1 & 1 \\
 0 & 1 & 0 & 1 & 1 & 0 & 1 & 0 & 1 & 1 \\
 1 & 0 & 0 & 0 & 0 & 0 & 1 & 0 & 0 & 1 \\
 0 & 1 & 0 & 0 & 0 & 0 & 0 & 1 & 1 & 0 \\
 0 & 0 & 1 & 0 & 1 & 0 & 0 & 0 & 0 & 1 \\
 0 & 0 & 0 & 1 & 0 & 1 & 0 & 0 & 1 & 0 \\
    \end{pmatrix}
\end{equation}
Notice this EBR matrix in Eq.~(\ref{eqn_EBRmat_3}) is identical to Eq.~(\ref{eqn_EBRmat}), although their bases have different meanings. 
We again use Eq.~(\ref{eqn_smith_nw}) to find the symmetry indicator formulas for $n_a$, $n_b$ and $n_c$ in Eqs.~(\ref{eqn_na_2}),~(\ref{eqn_nb_2}), and~(\ref{eqn_nc_2}).

\subsection{$p4/m$}

In the case of layer group $p4/m$, the site-symmetry groups of the maximal Wyckoff positions are defined in Table~\ref{tab_ssgroups} and their irreps are defined in Table ~\ref{tab_irrep_I_noT}. The basis of momentum space irreps is:
\begin{align}
\big(~
    &{}^1E_{\frac12g}^{\Gamma}, {}^1E_{\frac32g}^{\Gamma}, {}^2E_{\frac12g}^{\Gamma}, {}^2E_{\frac32g}^{\Gamma},
    {}^1E_{\frac12u}^{\Gamma}, {}^1E_{\frac32u}^{\Gamma}, {}^2E_{\frac12u}^{\Gamma}, ^2E_{\frac32u}^{\Gamma},\nonumber \\
    &{}^1E_{\frac12g}^{X}, {}^2E_{\frac12g}^{X}, {}^1E_{\frac12u}^{X}, {}^2E_{\frac12u}^{X}, \nonumber \\
    &{}^1E_{\frac12g}^{M}, {}^1E_{\frac32g}^{M}, {}^2E_{\frac12g}^{M}, {}^2E_{\frac32g}^{M}
    {}^1E_{\frac12u}^{M}, {}^1E_{\frac32u}^{M}, {}^2E_{\frac12u}^{M}, {}^2E_{\frac32u}^{M}
    ~\big)
\end{align}
The basis of EBRs is:
\begin{align}
\big(~
   &^1E_{\frac12g}^{1a}, ^1E_{\frac32g}^{1a}, ^2E_{\frac12g}^{1a}, ^2E_{\frac32g}^{1a},
    ^1E_{\frac12u}^{1a}, ^1E_{\frac32u}^{1a}, ^2E_{\frac12u}^{1a}, ^2E_{\frac32u}^{1a}, \nonumber\\
   &^1E_{\frac12g}^{1b}, ^1E_{\frac32g}^{1b}, ^2E_{\frac12g}^{1b}, ^2E_{\frac32g}^{1b},
    ^1E_{\frac12u}^{1b}, ^1E_{\frac32u}^{1b}, ^2E_{\frac12u}^{1b}, ^2E_{\frac32u}^{1b}, \nonumber\\
   &^1E_{\frac12g}^{2c}, ^2E_{\frac12g}^{2c},^1E_{\frac12u}^{2c}, ^2E_{\frac12u}^{2c}
~\big)
\end{align}
Then the EBR matrix is: $A=$
\begin{equation}
\small
    \label{eqn_EBRmat_4}
    \left(
\begin{array}{cccccccccccccccccccc}
 1 & 0 & 0 & 0 & 0 & 0 & 0 & 0 & 1 & 0 & 0 & 0 & 0 & 0 & 0 & 0 & 1 & 0 & 0 & 0 \\
 0 & 1 & 0 & 0 & 0 & 0 & 0 & 0 & 0 & 1 & 0 & 0 & 0 & 0 & 0 & 0 & 0 & 1 & 0 & 0 \\
 0 & 0 & 1 & 0 & 0 & 0 & 0 & 0 & 0 & 0 & 1 & 0 & 0 & 0 & 0 & 0 & 1 & 0 & 0 & 0 \\
 0 & 0 & 0 & 1 & 0 & 0 & 0 & 0 & 0 & 0 & 0 & 1 & 0 & 0 & 0 & 0 & 0 & 1 & 0 & 0 \\
 0 & 0 & 0 & 0 & 1 & 0 & 0 & 0 & 0 & 0 & 0 & 0 & 1 & 0 & 0 & 0 & 0 & 0 & 1 & 0 \\
 0 & 0 & 0 & 0 & 0 & 1 & 0 & 0 & 0 & 0 & 0 & 0 & 0 & 1 & 0 & 0 & 0 & 0 & 0 & 1 \\
 0 & 0 & 0 & 0 & 0 & 0 & 1 & 0 & 0 & 0 & 0 & 0 & 0 & 0 & 1 & 0 & 0 & 0 & 1 & 0 \\
 0 & 0 & 0 & 0 & 0 & 0 & 0 & 1 & 0 & 0 & 0 & 0 & 0 & 0 & 0 & 1 & 0 & 0 & 0 & 1 \\
 1 & 0 & 1 & 0 & 0 & 0 & 0 & 0 & 0 & 0 & 0 & 0 & 0 & 1 & 0 & 1 & 1 & 0 & 0 & 1 \\
 0 & 1 & 0 & 1 & 0 & 0 & 0 & 0 & 0 & 0 & 0 & 0 & 1 & 0 & 1 & 0 & 0 & 1 & 1 & 0 \\
 0 & 0 & 0 & 0 & 1 & 0 & 1 & 0 & 0 & 1 & 0 & 1 & 0 & 0 & 0 & 0 & 0 & 1 & 1 & 0 \\
 0 & 0 & 0 & 0 & 0 & 1 & 0 & 1 & 1 & 0 & 1 & 0 & 0 & 0 & 0 & 0 & 1 & 0 & 0 & 1 \\
 1 & 0 & 0 & 0 & 0 & 0 & 0 & 0 & 0 & 0 & 1 & 0 & 0 & 0 & 0 & 0 & 0 & 0 & 0 & 1 \\
 0 & 1 & 0 & 0 & 0 & 0 & 0 & 0 & 0 & 0 & 0 & 1 & 0 & 0 & 0 & 0 & 0 & 0 & 1 & 0 \\
 0 & 0 & 1 & 0 & 0 & 0 & 0 & 0 & 1 & 0 & 0 & 0 & 0 & 0 & 0 & 0 & 0 & 0 & 0 & 1 \\
 0 & 0 & 0 & 1 & 0 & 0 & 0 & 0 & 0 & 1 & 0 & 0 & 0 & 0 & 0 & 0 & 0 & 0 & 1 & 0 \\
 0 & 0 & 0 & 0 & 1 & 0 & 0 & 0 & 0 & 0 & 0 & 0 & 0 & 0 & 1 & 0 & 0 & 1 & 0 & 0 \\
 0 & 0 & 0 & 0 & 0 & 1 & 0 & 0 & 0 & 0 & 0 & 0 & 0 & 0 & 0 & 1 & 1 & 0 & 0 & 0 \\
 0 & 0 & 0 & 0 & 0 & 0 & 1 & 0 & 0 & 0 & 0 & 0 & 1 & 0 & 0 & 0 & 0 & 1 & 0 & 0 \\
 0 & 0 & 0 & 0 & 0 & 0 & 0 & 1 & 0 & 0 & 0 & 0 & 0 & 1 & 0 & 0 & 1 & 0 & 0 & 0 \\
\end{array}
\right)
\end{equation}

After Smith decomposition (Eq.~(\ref{eqn_smith})), the diagonal matrix is:
\begin{equation}
    D=\text{diag}(1,1,1,1,1,1,1,1,1,1,1,1,1,1,4,4,0,0,0,0)
\end{equation}
and $V=$
\begin{equation}
\left(\tiny
\begin{array}{cccccccccccccccccccc}
   1 & 0 & 0 & 0 & 0 & 0 & 0 & 0 & 0 & 0 & 0 & 0 & 0 & 0 & 1 & 0 & 0 & -1 & 0 & -1 \\
 0 & 1 & 0 & 0 & 0 & 0 & 0 & 0 & 0 & 0 & 0 & 0 & 0 & 0 & 0 & 1 & -1 & 0 & -1 & 0 \\
 0 & 0 & 1 & 0 & 0 & 0 & 0 & 0 & 0 & 0 & -1 & 0 & 0 & 0 & -1 & 0 & 0 & -1 & 0 & -1 \\
 0 & 0 & 0 & 1 & 0 & 0 & 0 & 0 & 0 & 0 & 0 & -1 & 0 & 0 & 0 & -1 & -1 & 0 & -1 & 0 \\
 0 & 0 & 0 & 0 & 1 & 0 & 0 & 0 & 0 & 0 & 0 & -1 & 0 & 0 & 0 & -2 & -1 & 0 & -1 & 0 \\
 0 & 0 & 0 & 0 & 0 & 1 & 0 & 0 & 0 & 0 & -1 & 0 & 0 & 0 & -2 & 0 & 0 & -1 & 0 & -1 \\
 0 & 0 & 0 & 0 & 0 & 0 & 1 & 0 & 0 & 0 & 0 & 0 & 0 & 0 & 0 & 0 & -1 & 0 & -1 & 0 \\
 0 & 0 & 0 & 0 & 0 & 0 & 0 & 1 & 0 & 0 & 0 & 0 & 0 & 0 & 0 & 0 & 0 & -1 & 0 & -1 \\
 0 & 0 & 0 & 0 & 0 & 0 & 0 & 0 & 1 & 0 & 0 & 0 & 0 & 0 & 1 & 0 & 0 & 1 & 0 & 0 \\
 0 & 0 & 0 & 0 & 0 & 0 & 0 & 0 & 0 & 1 & 0 & 0 & 0 & 0 & 0 & 1 & 1 & 0 & 0 & 0 \\
 0 & 0 & 0 & 0 & 0 & 0 & 0 & 0 & 0 & 0 & 0 & 0 & 0 & 0 & -1 & 0 & 0 & 1 & 0 & 0 \\
 0 & 0 & 0 & 0 & 0 & 0 & 0 & 0 & 0 & 0 & 0 & 0 & 0 & 0 & 0 & -1 & 1 & 0 & 0 & 0 \\
 0 & 0 & 0 & 0 & 0 & 0 & 0 & 0 & 0 & 0 & 0 & -1 & 1 & 0 & 0 & -2 & 1 & 0 & 0 & 0 \\
 0 & 0 & 0 & 0 & 0 & 0 & 0 & 0 & 0 & 0 & -1 & 0 & 0 & 1 & -2 & 0 & 0 & 1 & 0 & 0 \\
 0 & 0 & 0 & 0 & 0 & 0 & 0 & 0 & 0 & 0 & 0 & 0 & 0 & 0 & 0 & 0 & 1 & 0 & 0 & 0 \\
 0 & 0 & 0 & 0 & 0 & 0 & 0 & 0 & 0 & 0 & 0 & 0 & 0 & 0 & 0 & 0 & 0 & 1 & 0 & 0 \\
 0 & 0 & 0 & 0 & 0 & 0 & 0 & 0 & 0 & 0 & 1 & 0 & 0 & 0 & 2 & 0 & 0 & 0 & 0 & 1 \\
 0 & 0 & 0 & 0 & 0 & 0 & 0 & 0 & 0 & 0 & 0 & 1 & 0 & 0 & 0 & 2 & 0 & 0 & 1 & 0 \\
 0 & 0 & 0 & 0 & 0 & 0 & 0 & 0 & 0 & 0 & 0 & 0 & 0 & 0 & 0 & 0 & 0 & 0 & 1 & 0 \\
 0 & 0 & 0 & 0 & 0 & 0 & 0 & 0 & 0 & 0 & 0 & 0 & 0 & 0 & 0 & 0 & 0 & 0 & 0 & 1 \\
    \end{array}
    \right)
\end{equation}
The matrix $U$ is too unwieldy to present here.\par

We again use Eq.~(\ref{eqn_smith_nw}) to find the symmetry indicator formulas for $n_a$, $n_b$ and $n_c$ in
Eqs.~(\ref{eqn_na_3}),~(\ref{eqn_nb_3}) and~(\ref{eqn_nc_3}).

\subsection{$p41'$}
\label{sec_Smith_p41p}

In $p41'$, the site-symmetry groups of the maximal Wyckoff positions are listed in Table~\ref{tab_ssgroups}.
Table~\ref{tab_irrep} shows that their irreps are identical to those of $p4/m'$.
Thus, the EBR matrix and its Smith normal form is the same as for $p4/m'$ in Sec.~\ref{sec_Smith_p4mp} and
$n_a$ and $n_b$ are given by Eqs.~(\ref{eqn_na1}) and (\ref{eqn_nb1}), respectively, while $n_c$ is determined only modulo one.

\par

\section{\label{app_diamond}Filling anomaly when the boundary is rotated by $45^\circ$ relative to the unit cell }
\begin{figure}
    \centering
    \includegraphics[width=\linewidth]{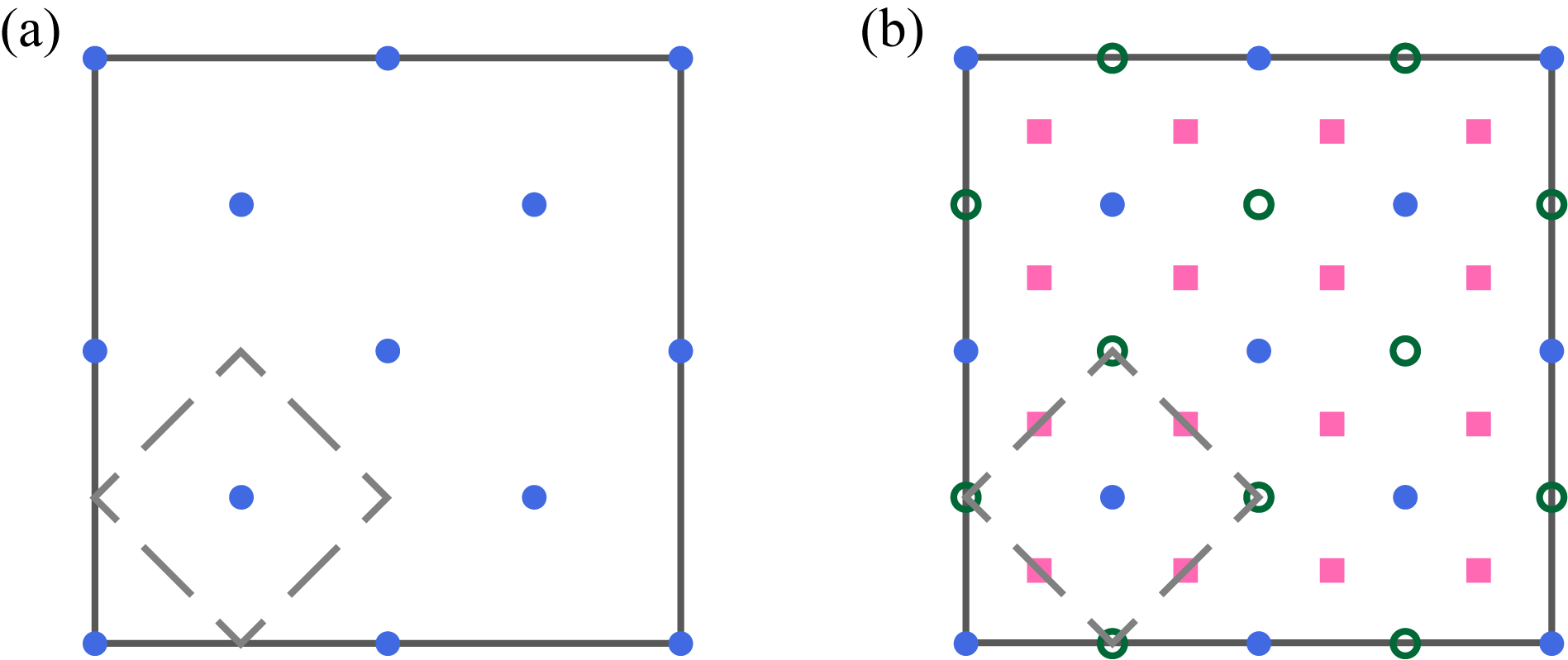}
    \caption{Square lattice whose boundary is rotated $45^\circ$ relative to the primitive unit cell, which is indicated by the dashed square. (a) Atoms occupy one Wyckoff position. (b) Atoms occupy multiple Wyckoff positions. }
    \label{fig_lattice_d}
\end{figure}

We derive an expression for the filling anomaly for a finite square lattice whose boundaries are rotated $45^\circ$ relative to the primitive unit cell, as shown in Fig.~\ref{fig_lattice_d}.
We consider the general case where there may be atoms at any number of Wyckoff positions (shown in Fig.~\ref{fig_lattice_d}(b));
the case where there is only one Wyckoff position occupied by atoms (shown in Fig.~\ref{fig_lattice_d}(a)) is a special case.

The Wyckoff positions are defined with respect to the primitive unit cell. 
To compute the filling anomaly, we need only consider the maximal Wyckoff positions, as we argued in Sec.~\ref{sec_2d_method}. The donated electrons from each maximal Wyckoff position are denoted by $a_a$, $a_b$, and $a_c$. The number of filled bands is $N=a_a+a_b+2a_c$.\par
The number of atoms, $N_w(L)$, at each Wyckoff position $w$ in this terminated square lattice with $L$ atoms along each side is given by
\begin{align}
\label{eqn_N_w_bd}
    N_{a}(L)&=(L-1)^2+L^2 \nonumber\\
    N_b(L)&= 2L(L-1) \nonumber \\
    N_c(L) &= 4(L-1)^2
\end{align}
(Notice these are different than in Eq.~(\ref{eqn_N_w_ac}), where $N_w(L)$ is computed with the boundary parallel to the unit cell.)
From Eqs.~(\ref{eqn_eta_square}) and~(\ref{eqn_N_w_bd}), we find the filling anomaly:
\begin{align}
    \eta = &2(N-dn_a-dn_b-2d n_c)L^2 \nonumber \\
    &-2(a_a+a_b+4a_c-dn_a-dn_b-4dn_c)L \nonumber \\
    &+(a_a-dn_a+4a_c-4dn_c)\qquad ~\text{mod}~4(\text{or}~8).
    \label{eqn_eta_11normal}
\end{align}
The bulk charge is determined by the number scaling with $L^2$ in this expression and must be zero:
\begin{equation}
\label{eqn_diamond_zero_charge}
    2(a_a-dn_a)+2(a_b-dn_b)+4(a_c-dn_c) = 0\mod 4(\text{or}~8),
\end{equation}
where we have used the expression for $N$ in Eq.~(\ref{eqn_defN}).
The bulk polarization parallel to the boundaries
is determined by the coefficient of $L$.
Since we are interested in polarization-free systems, the term proportional to $L$ must vanish:
\begin{equation}
\label{eqn_diamond_zero_polarization}
    2(a_a-dn_a)+2(a_b-dn_b) +8(a_c-dn_c) = 0\mod 4(\text{or}~8)
\end{equation}
The last term in Eq.~(\ref{eqn_diamond_zero_polarization}) vanishes mod 4 or mod 8.
In addition, the last term in Eq.~(\ref{eqn_diamond_zero_charge}) clearly vanishes mod 4, but also vanishes when taken mod 8, which applies in the presence of time-reversal symmetry, because time-reversal requires that $a_c$ be even and $d=2$ ($d$ is defined in Eq.~(\ref{eqn_defd}).)
Thus, the only constraint from Eqs.~(\ref{eqn_diamond_zero_charge}) and~(\ref{eqn_diamond_zero_polarization}) is 
\begin{equation}
    \label{eqn_diamond_zero_constraint}
    2(a_a-dn_a)+2(a_b-dn_b) = 0\mod 4(\text{or}~8)
\end{equation}
{Notice that this constraint differs from the constraint in Eq.~(\ref{eqn_pol0}), which applies when the boundary is parallel to the unit cell. As a result $[X_u]$ need not be even, as we found in Table~\ref{tab_BCT}.}
\par
The filling anomaly $\eta$ is determined by the $L$-independent term:
\begin{equation}
\label{eqn_eta_na_d}
    \eta = a_a-dn_a ~\mod 4(\text{or}~8),
\end{equation}
where we have simplified the $L$-independent term from Eq.~(\ref{eqn_eta_11normal}) by taking it mod 4 (or mod 8).
Notice Eq.~(\ref{eqn_eta_na_d}) is identical to Eq.~(\ref{eqn_eta_na}), which is derived for the case where the boundary is parallel to the unit cell.
Therefore, the symmetry indicators will be the same in the two cases.

\section{\label{app_J} Spin-3/2 matrices}
Here we explicitly define the spin-3/2 matrices that are used in Eq.~(\ref{eqn_C4}) and (\ref{eqn_TR}):

\begin{align}
    J_{x} &= \left(\begin{array}{cccc}{0} & {\frac{\sqrt{3}}{2}} & {0} & {0} \\ {\frac{\sqrt{3}}{2}} & {0} & {1} & {0} \\ 
    {0} & {1} & {0} & {\frac{\sqrt{3}}{2}} \\ 
    {0} & {0} & {\frac{\sqrt{3}}{2}} & {0}\end{array}\right)
\\
    J_{y} &= \left(\begin{array}{cccc}{0} & {-i \frac{\sqrt{3}}{2}} & {0} & {0} \\ {i \frac{\sqrt{3}}{2}} & {0} & {-i} & {0} \\ {0} & {i} & {0} & {-i \frac{\sqrt{3}}{2}} \\ {0} & {0} & {i \frac{\sqrt{3}}{2}} & {0}\end{array}\right)
\\
    J_{z} &= \left(\begin{array}{cccc}{\frac{3}{2}} & {0} & {0} & {0} \\ {0} & {\frac{1}{2}} & {0} & {0} \\ {0} & {0} & {-\frac{1}{2}} & {0} \\ {0} & {0} & {0} & {-\frac{3}{2}}\end{array}\right)
\end{align}

\section{\label{app_model}Details of BCT tight-binding model}

In this appendix, we provide additional details about the BCT model studied in Sec.~\ref{sec_3d}.
The BCT model has the symmetry of space group $87$ $I4/m$, as well as time-reversal symmetry $\cal T$. 

The unit cell and hoppings are illustrated in Figs.~\ref{fig_3d}(a) and (b). The model includes hopping to the four nearest and eight next-nearest (if $c>a$) atoms. The relation between the two hopping terms shown in Eq.~(\ref{eqn_model}) and the other ten hopping terms are:
\begin{align}
    V_{0\rightarrow e_1-e_z}&= (C_2{\cal I})^{-1} V_{0\rightarrow e_1} C_2{\cal I} \nonumber\\
    V_{0\rightarrow e_2}&= C_4^{-1}V_{0\rightarrow e_1}C_4\nonumber \\
    V_{0\rightarrow e_2-e_z}&= C_4{\cal I}V_{0\rightarrow e_1} (C_4{\cal I})^{-1}\nonumber \\
    V_{0\rightarrow e_3-e_y}&= C_4^{-1} V_{0\rightarrow e_3} C_4,
    \label{eqn_relation}
\end{align}
and
\begin{equation}
\label{eqn_relation_r}
    V_{0\rightarrow \mathbf{r}} = V^{\dagger}_{0\rightarrow -\mathbf{r}},
\end{equation}
where the matrix forms of the symmetry generators are given in Eqs.~(\ref{eqn_C4}), (\ref{eqn_Inv}) and (\ref{eqn_TR}), and $\mathcal{C}_2 \equiv \mathcal{C}_4^2$.

The primitive lattice has one site in the unit cell. 
Since each site has a spin-$\frac{3}{2}$ degree of freedom, the Hamiltonian, defined by Eqs.~(\ref{eqn_model}), (\ref{eqn_relation}), and (\ref{eqn_relation_r}) is a $4\times 4$ matrix, given by:
\begin{align}
    \label{eqn_ham_p}
    H^{p} =& t(\cos{\frac{k_x}{2}}\cos{\frac{k_z}{2}}+\cos{\frac{k_y}{2}}\cos{\frac{k_z}{2}}-2\cos{\frac{k_x}{2}}\cos{\frac{k_y}{2}}) \tau_z\sigma_z \nonumber\\
    &+ \gamma \sin{\frac{k_z}{2}}\tau_z(\sin{\frac{k_x}{2}}\sigma_x+\sin{\frac{k_y}{2}}\sigma_y) \nonumber\\
    &+\gamma\sin{\frac{k_x}{2}}\sin{\frac{k_y}{2}}\tau_y\sigma_0 \nonumber\\
    &+\beta (\cos{\frac{k_y}{2}}-\cos{\frac{k_x}{2}})\cos{\frac{k_z}{2}}\tau_+\sigma_0 \nonumber \\
    &+\beta^* (\cos{\frac{k_y}{2}}-\cos{\frac{k_x}{2}})\cos{\frac{k_z}{2}}\tau_-\sigma_0 \nonumber\\
    &+m\tau_z\sigma_z,
\end{align}
where $k_x$, $k_y$, $k_z$ correspond to the momenta reciprocal to $e_x$, $e_y$, $e_z$.\par

\subsection{Hamiltonian and symmetry operators in conventional unit cell}

The conventional unit cell contains two sublattices, as indicated in Fig.~\ref{fig_3d}(a).
Since each site hosts a spin-$\frac{3}{2}$ degree of freedom, 
the Hamiltonian in the conventional unit cell in reciprocal space is an $8\times 8$ matrix, which takes the form
\begin{align}
    \label{eqn_ham_c}
    H^{c} = &(m-2t\cos{\frac{k_x}{2}}\cos{\frac{k_y}{2}})\rho_0\tau_z\sigma_z \nonumber\\
    &+\gamma \sin{\frac{k_x}{2}}\sin{\frac{k_y}{2}}\rho_0\tau_y\sigma_0 \nonumber\\
    &+t(\cos{\frac{k_x}{2}}+\cos{\frac{k_y}{2}})\cos{\frac{k_z}{2}}\rho_x\tau_z\sigma_z \nonumber\\
    &+\gamma \sin{\frac{k_z}{2}}\rho_x\tau_z(\sin{\frac{k_x}{2}}\sigma_x+\sin{\frac{k_y}{2}}\sigma_y) \nonumber \\
    &+\beta(\cos{\frac{k_y}{2}}-\cos{\frac{k_x}{2}})\cos{\frac{k_z}{2}} \rho_x \tau_+\sigma_0 \nonumber \\
    &+\beta^* (\cos{\frac{k_y}{2}}-\cos{\frac{k_x}{2}})\cos{\frac{k_z}{2}}\rho_x\tau_-\sigma_0, 
\end{align}
where again $k_x$, $k_y$, $k_z$ correspond to the basis reciprocal to $e_x$, $e_y$, $e_z$,
and we have introduced an additional set of Pauli matrices $\rho_i$ to indicate the sublattice degree of freedom.
As in the main text and in the Hamiltonian in Eq.~(\ref{eqn_ham_p}), $\sigma_i$ and $\tau_i$ act on the hybrid spin and orbital degrees of freedom.

In this basis, the symmetry operators are implemented by the matrices:
\begin{align}
C^{c}_4 &= \begin{pmatrix}
C_4 & \\
& C_4
\end{pmatrix} \nonumber \\
{\cal I}^{c} & = \begin{pmatrix}
{\cal I} & \\
& {\cal I}
\end{pmatrix}
    \label{eqn_sym_c}
\end{align}
where the superscript $c$ indicates the conventional unit cell basis. $C_4$ and $\cal I$ are defined for the primitive lattice in Eqs.~(\ref{eqn_C4}) and (\ref{eqn_Inv}).\par


\subsection{\label{app_model_para}Parameters for numerical calculations}
There is an unexpected anti-unitary symmetry in the $k_z=\pi$ plane. This symmetry is artificial because it can be broken by adding small next-next-nearest hopping terms that preserve all the symmetries. The extra terms do not change the topology and do not break any symmetry, but will help to eliminate unphysical gapless surface states which are protected by the artificial symmetry. These small next-nearest hopping terms are:
\begin{align}
    V_{0\rightarrow e_x} &= b(\tau_z\sigma_z-\sqrt{3}\tau_x\sigma_0), \nonumber\\
    V_{0\rightarrow e_y} &= b(\tau_z\sigma_z+\sqrt{3}\tau_x\sigma_0), \nonumber \\
    V_{0\rightarrow e_z} &= -2b\tau_z\sigma_z.
\end{align}

We consider a system that is finite in the $x$ and $y$ directions, i.e., its boundaries are normal to $e_x$ and $e_y$, as shown in Fig.~\ref{fig_3d}(b).
We terminate the boundary to respect $C_4$ symmetry.

In our numerical calculations, the parameters are set to be: $t=1$, $\beta=1+i$, $\gamma=1$, $b=0.2$. The parameters are taken in such a way that when $m=0$, there is a quadratic band touching at $\Gamma$. 
The rod states are calculated for a square of size $15 e_x$ by $15 e_y$. In Fig.~\ref{fig_3d}(c), $m=-0.5$, realizing the HOTI phase. In Fig.~\ref{fig_3d}(d), $m=1$, realizing the DSM(i) phase. \par

The phase diagram in Fig.~\ref{fig_3d}(e) is derived with $b=0$. This small value of $b=.2$ only slightly changes the phase transition points of $m/t$.
\par

\subsection{Symmetry eigenvalues in the HOTI phase and DSM(i) phase}
\label{sec_3d_irreps}

\begin{table}[h]
\begin{tabular}{|c|cccc|}
\hline
HSP&$E_{\frac12 g}$&$E_{\frac12 u}$&$E_{\frac32 g}$&$E_{\frac32 u}$\\
\hline
$\Gamma$&2&0&0&0\\
$X$&1&1&0&0\\
$M$&1&0&1&0\\
$Z$&1&1&0&0\\
$T$&2&0&0&0\\
$R$&0&1&1&0\\
\hline
\end{tabular}
\caption{Symmetry eigenvalues at high symmetry points in the HOTI phase. $\Gamma$, $X$, $M$ reside in the $k_z=0$ plane, while $Z$, $T$, $R$ reside in the $k_z=\pi$ plane. The filling anomaly for each plane is calculated in Table~\ref{tab_BCT}.}
\label{tab_calculation_HOTI}
\end{table}
\par

\begin{table}[h]
\begin{tabular}{|c|cccc|cc|}
\hline
HSP&$E_{\frac12 g}$&$E_{\frac12 u}$&$E_{\frac32 g}$&$E_{\frac32 u}$ &$E_{\frac12}$&$E_{\frac32}$\\
\hline
$\Gamma$&1&0&1&0 &1&1\\
$X$&1&1&0&0 &2&0\\
$M$&1&0&1&0 &1&1\\
\hline
\hline
$Z$&2&0&0&0 &2&0\\
$T$&1&1&0&0 &2&0\\
$R$&0&1&1&0 &1&1\\
\hline
\end{tabular}
\caption{Symmetry eigenvalues at high symmetry points in the DSM(i) phase. $\Gamma$, $X$, $M$ reside in the $k_z=0$ plane, while $Z$, $T$, $R$ reside in the $k_z=\pi$ plane. The $k_z=0$ plane has mirror Chern number $C_m=2$, while the $k_z=\pi$ plane is in a fragile topological phase and has filling anomaly $\eta=4$. There is a bulk Dirac point in the plane $k_z=k_0$. At $k_z$ planes between $0$ and $k_0$, the symmetry eigenvalues correspond to the last two columns $E_{\frac12}$ and $E_{\frac32}$ for $\Gamma$, $X$, $M$ as required by the conservation of angular momentum (or equivalently the compatibility relation). At $k_z$ planes between $k_0$ and $\pi$, the symmetry eigenvalues correspond to the last two columns $E_{\frac12}$ and $E_{\frac32}$ for $Z$, $T$, $R$.}
\label{tab_calculation_DSMi}
\end{table}

We list the the symmetry eigenvalues computed from our model in the HOTI phase in Table~\ref{tab_calculation_HOTI}. The results are used to calculate the filling anomaly in Table~\ref{tab_BCT}.\par

The symmetry eigenvalues computed in the DSM(i) phase are listed in Table~\ref{tab_calculation_DSMi}. The results are used to calculate the filling anomaly in Table~\ref{tab_BCT_HOFA}. 
In addition, the $k_z=0$ plane has mirror Chern number $C_m=2$. The mirror Chern number can be evaluated by calculating the Chern numbers of the $+i$ and $-i$ sectors. The symmetry indicator formula for the Chern number is~\cite{fang2012bulk}:
\begin{equation}
\label{eqn_chern_number}
    i^{C}=\prod_{i \in occ.}(-1)^{F} \xi_{i}(\Gamma) \xi_{i}(M) \zeta_{i}(Y)
\end{equation}
where $F$ is twice the total spin and can be replaced with the number of filled bands $N$ in our spinful case; $\xi_i$ is the $C_4$ eigenvalue of the $i^{\rm th}$ band; and $\zeta_i$ is the $C_2$ eigenvalue of the $i^{\rm th}$ band. Time-reversal symmetry ensures that the number of filled band in each mirror sector is $N/2$. Time-reversal symmetry also constrains the mirror Chern number: $C_m=\frac12(C_{+i}
+C_{-i})=C_{+i}$ . 
{We can determine $C_{+i}$ by counting the numbers of irreps that contain mirror eigenvalue $+i$ and plug them into Eq.~(\ref{eqn_chern_number}). For example, $E^{X}_{\frac12 u}$ has two components: $\zeta(X)=i$ and $\zeta(X)=-i$. Since this irrep has inversion eigenvalue ${\cal I}=-1$, 
only the $C_2$ eigenvalue $\zeta(X)=-i$ corresponds to the sector with mirror eigenvalue $+i$.
Thus, $(-i)^{\# E^X_{\frac12 u}}=(i)^{-\# E^X_{\frac12 u}}$ in Eq.~(\ref{eqn_chern_number}). The other irreps come into the equation similarly.}
From these facts, we obtain the symmetry indicator formula for the mirror Chern number for our layer group $p4/m1'$ at this $k_z=0$ plane ($C_4$, $\cal T$ and $\cal I$):
\begin{align}
    \label{eqn_mirror_chern}
    C_m = N&+\left(\# E^X_{\frac12 g}- \# E^X_{\frac12 u}\right) +\frac12 \sum_{i=\Gamma,M} \left(\# E^i_{\frac12 g}- \# E^i_{\frac12 u}\right) \nonumber \\
    &-\frac32 \sum_{i=\Gamma,M}\left(\# E^i_{\frac32 g}- \# E^i_{\frac32 u}\right)\quad \text{mod}~4
\end{align}
Plugging in the symmetry eigenvalues from Table~\ref{tab_calculation_HOTI}, we find $C_m=2$ at $k_z=0$, and $C_m=0$ at $k_z=\pi$. For the HOTI phase, the mirror Chern number $C_m=0$ for both $k_z=0$ and $k_z=\pi$ planes, as we can verify by plugging the symmetry eigenvalues from Table~\ref{tab_calculation_DSMi} into Eq.~(\ref{eqn_mirror_chern}). 
\par
There are a pair of bulk Dirac points at $(0,0,\pm k_0)$. The symmetry eigenvalues at a $k$-plane between $0$ and $k_0$ can be obtained by "forgetting" the inversion eigenvalues of $\Gamma$, $X$ and $M$. The symmetry eigenvalues at a $k$-plane between $k_0$ and $\pi$ can be obtained by "forgetting" the inversion eigenvalues of $Z$, $T$ and $R$.
\par

\subsection{\label{app_model_integer}Numerical calculation for a rod geometry with integer number of unit cells}

\begin{figure}[b]
    \centering
    \includegraphics[width=.6\linewidth]{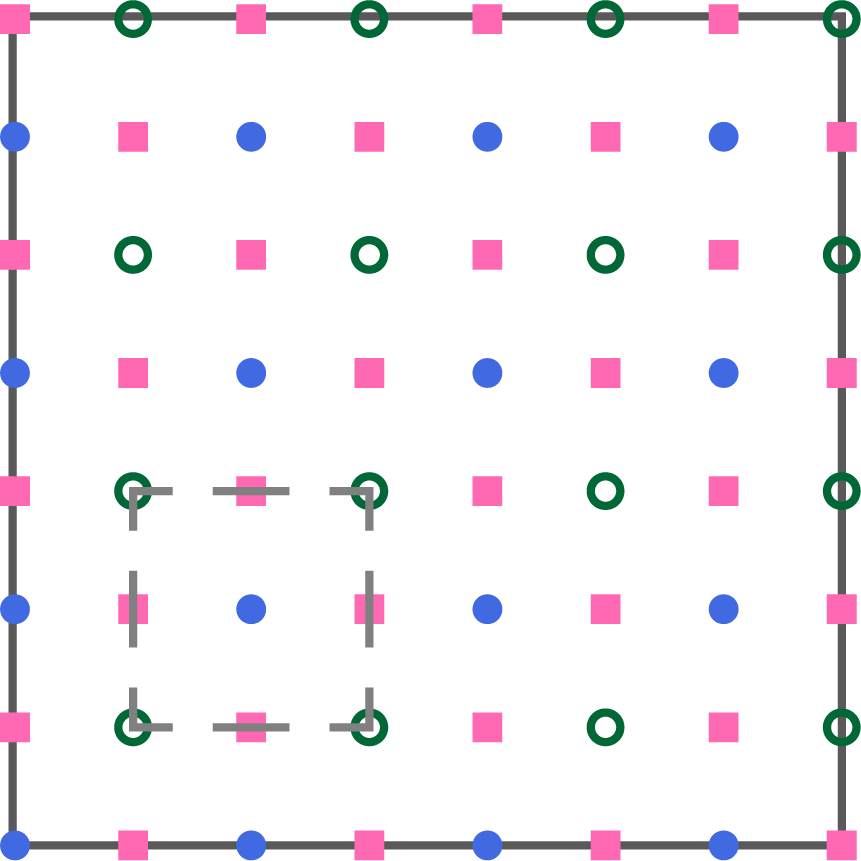}
    \caption{A square lattice terminated with an integer number of unit cells is not globally $C_4$-symmetric, i.e., if the crystal is rotated about a bulk $C_4$ center (blue or green dot), the rotated lattice does not coincide with the original lattice. This lattice should be compared to  Fig.~\ref{fig_lattice}(b), which shows a $C_4$-symmetric termination and has a fractional number of unit cells. Solid blue and hollow green dots and pink squares indicate atoms at the Wyckoff positions $1a$, $1b$ and $2c$, respectively. Dashed gray lines indicate the primitive unit cell.}
    \label{fig_lattice_app}
\end{figure}

The corner states are protected by the global $C_4$ symmetry.
Thus, rigorously, a $C_4$-symmetric termination of the lattice is required to protect the corner states.
However, in practice we find that the corner states survive on a lattice termination that is not globally $C_4$-symmetric if it has $C_4$ symmetry in the bulk (see Fig.~\ref{fig_lattice_app} for an example of such a lattice termination).
Physically, this is reasonable because if the corners are far apart, the local spectrum at one corner should not depend on the termination at another corner.



\begin{figure*}
    \centering
    \includegraphics[width=0.8\linewidth]{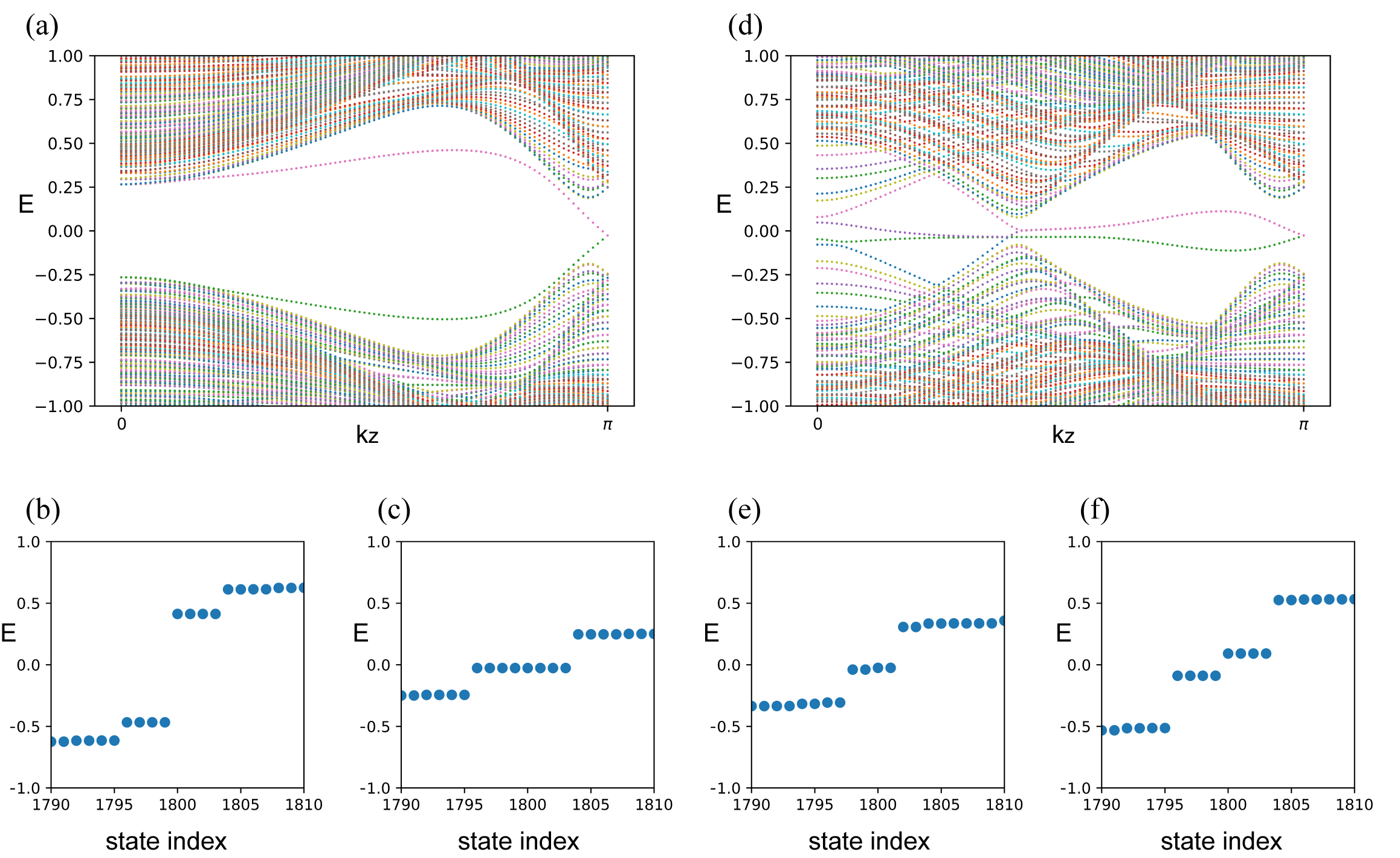}
    \caption{Spectrum for a rod (finite in the $x$- and $y$- directions, infinite in $z$) in (a) the HOTI phase and in (d) the DSM(i) phase. Every $k_z$ slice in this rod has an integer number ($15\times 15$) of unit cells, breaking the global $C_4$ symmetry. The number of electrons at charge neutrality is $1800$. The energy of each state near $E=0$
    is plotted for the following $k_z$ slices in the HOTI and DSM(i) phases: (b) HOTI phase, $k_z=\pi/2$, $\eta = 0 \mod 4$ 
    (c) HOTI phase, $k_z=\pi$, $\eta = 4\mod 8$; (e) DSM(i) phase, $k_z=\pi/4$, $\eta = 2 \mod 4$; (f) DSM(i) phase, $k_z=3\pi/4$, $\eta = 0 \mod 4$. 
    }
    \label{fig_integer}
\end{figure*}

In Fig.~\ref{fig_integer}, we numerically compute the rod states of our model with the same parameters as in App.~\ref{app_model_para} ($t=1$, $\beta=1+i$, $\gamma=1$, $b=0.2$), but with an integer number of unit cells, breaking the global $C_4$ symmetry. The spectrum is similar to Fig.~\ref{fig_3d}(c) and (d) (where global $C_4$ symmetry is preserved) for the same system size. Most importantly, the filling anomalies of each $k_z$ slice (obtained by counting the number of occupied bands) for the two choices of terminations are the same. 

\end{appendix}

\end{document}